\documentclass[onecolumn]{els-mrw} 
\usepackage[utf8]{inputenc}
\usepackage{amsmath,amssymb,amsfonts,amsthm,makeidx,graphicx}
\usepackage{txfonts}
\usepackage{helvet}
\usepackage{color}
\usepackage{hyperref}
\newcommand{\hsig}{H_{\Sigma}}
\newcommand{\hW}{\hat{W}}
\newcommand{\hV}{\hat{V}}
\newcommand{\tr}{\text{tr}}

\begin{document}

\chapter{Quantum analogues of exponential sensitivity:
from Loschmidt echo to Krylov complexity}\label{chapW}

\author[1]{Ignacio Garc\'ia-Mata}
\author[2]{Diego A. Wisniacki}
\address[1]{\orgname{Instituto de Investigaciones F\'isicas de Mar del Plata}, \orgdiv{CONICET \& Universidad Nacional de Mar del Plata}, \orgaddress{Funes 3350, 7600, Mar del Plata, Argentina.}}
\address[2]{\orgname{Departamento de F\'isica ``J. J. Giambiagi'' and IFIBA}, \orgdiv{FCEyN,
Universidad de Buenos Aires}, \orgaddress{Ciudad Universitaria, 1428 Buenos Aires, Argentina}}

\maketitle

\begin{glossary}[Keywords]
Quantum Chaos, Nonlinear Systems
\end{glossary}

\begin{abstract}[Abstract]
One of the fundamental manifestations of classical chaos is exponential sensitivity to initial conditions that is, two trajectories starting from nearly identical initial states diverge exponentially over time. This behavior is quantified by the Lyapunov exponents. Due to the unitary nature of quantum mechanics, such exponential divergence is elusive in quantum systems. As a result, several alternative quantities have been proposed and studied in recent years to capture analogous behavior.

In this article, we present a pedagogical overview of three such quantities that have been the focus of intense research in recent years: the Loschmidt echo, out-of-time-order correlators (OTOCs), and Krylov complexity.
\end{abstract}

\section{Introduction: Searching for sensitivity to initial conditions in quantum mechanics}

One of the defining features of classical chaos is the exponential sensitivity of trajectories to initial conditions, as quantified by Lyapunov exponents. This exponential divergence is at the heart of unpredictability in nonlinear dynamical systems. In quantum mechanics, however, the unitary nature of time evolution precludes a straightforward analogue: two quantum states evolving under slightly different Hamiltonians remain at fixed Hilbert-space distance. This apparent absence of sensitivity to initial conditions has motivated the search for suitable quantum indicators of instability and chaos.  

Several approaches have been proposed to capture quantum signatures of classical chaos. Early developments focused on \emph{spectral statistics}: chaotic quantum systems display level correlations described by random matrix theory, while integrable ones follow Poisson statistics~\cite{bohigas,haake,guhr1998random}. Beyond spectral measures, dynamical probes have emerged as particularly powerful. Among them, three quantities have gained central importance in recent decades: the \emph{Loschmidt echo}, \emph{out-of-time-ordered correlators (OTOCs)}, and \emph{Krylov complexity}. Each of these captures complementary aspects of information spreading, reversibility, and complexity in quantum dynamics.  

The \emph{Loschmidt echo} (or quantum fidelity), originally introduced by Peres~\cite{Peres1984} and later developed by Jalabert and Pastawski~\cite{Jalabert2001}, quantifies the overlap between an initial state and the state evolved under slightly perturbed dynamics. Its decay regimes reveal profound connections with classical chaos: in particular, the exponential decay governed by the Lyapunov exponent provides one of the clearest manifestations of quantum--classical correspondence~\cite{Gorin2006,Jacquod2009,DiegoScholar}. The Loschmidt echo has also become an important tool for studying decoherence, irreversibility, and quantum information stability~\cite{Prosen2002,Cucc2003,Quan2006}.  

The \emph{out-of-time-ordered correlator (OTOC)}, first appearing in the context of superconductivity~\cite{larkin1969quasiclassical} and later popularized in the study of quantum chaos, quantum information and holography~\cite{shenker2014black,maldacena2016, swingle2018, otocScholarpedia}, provides a dynamical measure of operator growth and information scrambling. In chaotic systems, OTOCs typically exhibit an early exponential growth characterized by a ``quantum Lyapunov exponent''~\cite{kitaev2015simple,otocScholarpedia}, directly analogous to classical instability. Their study has established deep connections between quantum chaos, thermalization, and high-energy physics, including the celebrated bound on chaos~\cite{maldacena2016}.  

More recently, the notion of \emph{Krylov complexity} has been introduced as a new and fresh perspective on quantum dynamics~\cite{Parker,nandy2025quantum,rabinovici2025krylov}. Built on the Lanczos recursion method, it quantifies the effective size of the Krylov subspace required to represent the time evolution of a state or operator. Krylov complexity has revealed new aspects of quantum chaos and integrability, particularly through the structure of Lanczos coefficients and their connection to random matrix behavior. This approach provides a unifying framework that links quantum sensitivity, operator spreading, and complexity growth in quantum systems.  

The purpose of this review is not to provide an exhaustive survey of the vast literature surrounding these topics---comprehensive reviews already exist~\cite{Gorin2006,Jacquod2009,DiegoScholar,nandy2025quantum,rabinovici2025krylov}---but rather to present a conceptual and pedagogical overview. By discussing the Loschmidt echo, OTOCs, and Krylov complexity within a unified framework, we aim to illuminate their historical origins, theoretical underpinnings, and interconnections, offering the reader a broad yet accessible entry point into the study of quantum analogues of classical exponential sensitivity.  

\section{The Loschmidt Echo: Fundamental Aspects of Quantum Reversibility}
The Loschmidt echo stands as one of the most revealing measures of quantum reversibility, providing deep insights into the stability of quantum systems under time-reversal operations \cite{goirin_PhysRep2006,Jacquod2009,DiegoScholar}. This review presents a thorough examination of the theoretical foundations underlying this phenomenon, beginning with its historical roots in the reversibility paradox and progressing to modern interpretations in quantum chaos theory. We analyze the characteristic decay regimes observed across different systems, from quantum chaotic to many-body configurations, while discussing the diverse experimental implementations that have shaped our current understanding. The discussion extends to the profound implications for quantum information science, decoherence studies, and fundamental questions in quantum mechanics.
%%%%%%%%%%%%%%%%%%%%%
\begin{figure}
\begin{center}
\includegraphics[width=0.95\linewidth]{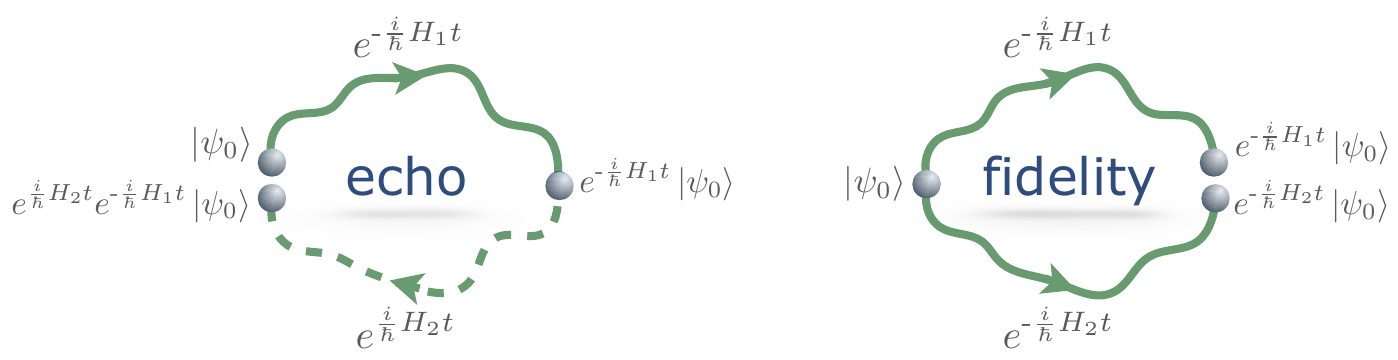}
\end{center}
\caption{Diagram showing the time-evolution process for (a) the Loschmidt echo as a measure of irreversibility and (b) the fidelity as a measure of sensitivity to perturbations.\label{fig:echofid}}
\end{figure}
%%%%%%%%%%%%%%%%%%%%%%
\subsection{Historical Foundations and Conceptual Framework}

The intellectual journey of the Loschmidt echo begins with one of the most profound debates in theoretical physics: the apparent contradiction between microscopic reversibility and macroscopic irreversibility. In the late 19th century, Joseph Loschmidt's keen observation about the time-reversal symmetry inherent in Newton's equations posed a significant challenge to Boltzmann's statistical interpretation of thermodynamics. This paradox, questioning how irreversible macroscopic behavior could emerge from reversible microscopic laws, planted the seeds for what would eventually evolve into the modern understanding of quantum reversibility through the Loschmidt echo \cite{wang2021microscope}.
%%%%%%%%%%%%%%%%%%%%%
\begin{figure}
\begin{center}
\includegraphics[width=0.95\linewidth]{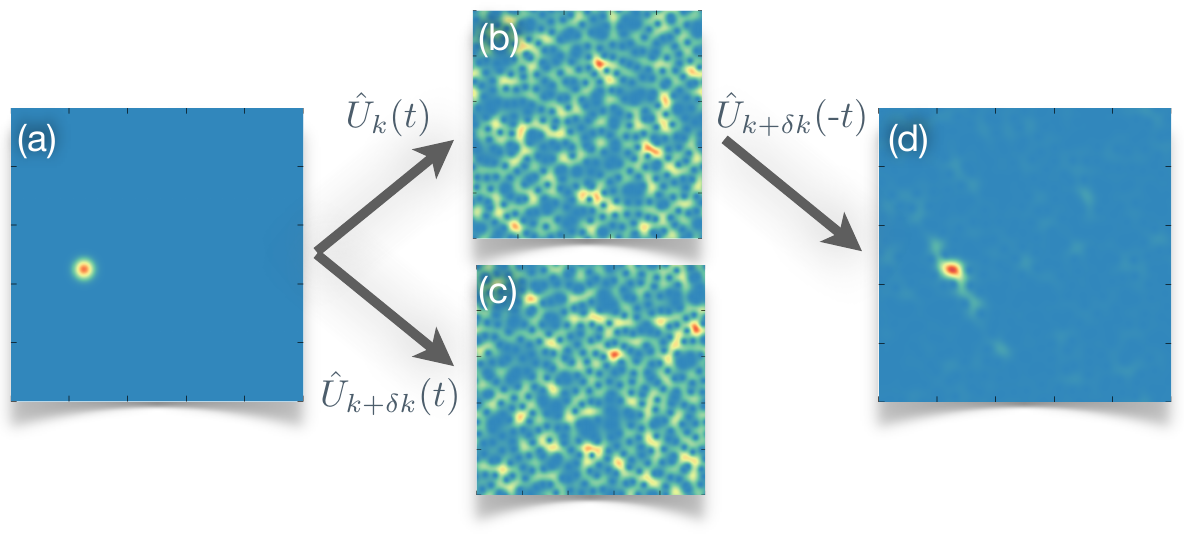}
\end{center}
\caption{Husimi representation of the wave-packet evolution in the perturbed Cat map \cite{dematos1995quantization} 
(a) Initial state at $t=0$. 
(b,c) Evolution with $\hat{U}_{k}$ and $\hat{U}_{k+\delta k}$ up to time $t$. 
(d) Backward evolution of (b) with $\dagger{\hat{U}_{k+\delta k}}$ during $(t,2t)$. 
The Loschmidt echo, given by the squared overlap between (a) and (d), is $M(t)=0.34$, equal to the {(modulus of the squared)} overlap between (b) and (c).
\label{fig:echogato}}
\end{figure}
%%%%%%%%%%%%%%%%%%%%%%

As quantum mechanics matured in the 20th century, these foundational questions found new expression. The seminal work of A. Peres in 1984 \cite{Peres1984} marked a turning point by introducing what we now recognize as the Loschmidt echo as a quantitative measure of quantum stability. However, it was the contribution of R. Jalabert and H. Pastawski in 2001 \cite{Jalabert2001} that truly established the modern theoretical framework, demonstrating the crucial connection between echo decay and classical Lyapunov exponents in chaotic systems.

Mathematically, the Loschmidt echo quantifies the overlap between an initial quantum state and its time-evolved counterpart after forward propagation under Hamiltonian $H_1$ followed by reversed propagation under Hamiltonian $H_2$ (see Fig. \ref{fig:echofid}). The perfect revival case occurs when $H_2$ exactly matches $H_1$, but in physical systems, unavoidable perturbations make this ideal situation unattainable.
The LE protocol involves three steps: (i) an initial state $|\psi_0\rangle$ evolves forward in time for a duration $t$ under a Hamiltonian $H_1$; (ii) the time-direction is effectively reversed, and the system evolves backward for the same duration $t$ under a Hamiltonian $H_2$; (iii) the overlap between the resulting state and the original state $|\psi_0\rangle$ is measured. 
The LE is defined as the squared magnitude of this overlap,
\begin{equation}
M(t) = \left| \langle \psi_0 | \, \exp(i H_2 t / \hbar) \, \exp(-i H_1 t / \hbar) | \psi_0 \rangle \right|^2.
\label{eq:LoschmidtEcho}
\end{equation}
In Fig. \ref{fig:echogato}, we show an example of the behavior of the Loschmidt echo in the perturbed cat map. We observe the evolution of the system under two slightly perturbed versions of the cat map. Although the two evolutions do not differ significantly, the value of the echo indicates that the time reversal is not effective. 

The backward evolution under $H_2$ is equivalent to a forward evolution under $-H_2$, embodying the quantum mechanical operation of time-reversal. Perfect recovery of the initial state ($M(t)=1$) is only possible if the reversal is exact, i.e., if $H_2 = H_1$. In practice, the reversal is always imperfect, and the Hamiltonians differ by a perturbation, $H_2 = H_1 + \kappa\Sigma$. Consequently, the LE decays from its initial value of unity. 
The resulting decay profile contains a wealth of information about the system's dynamics, irreversibility  and its sensitivity to perturbations. The rate and functional form of the decay --be it Gaussian, exponential, or algebraic -- depend critically on the strength $\kappa$ of the perturbation and the underlying classical dynamics of the system, making the LE a powerful tool for studying quantum chaos, decoherence, and the quantum-classical correspondence.
%%%%%%%%%%%%%%%%%%%%%
\begin{figure}
\begin{center}
\includegraphics[width=0.9\linewidth]{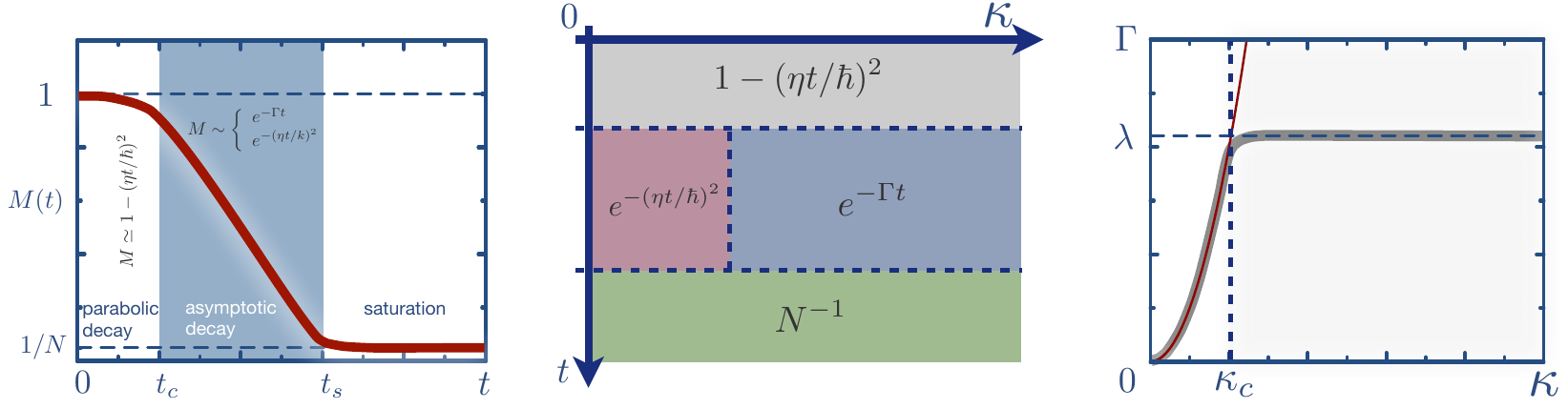}
\end{center}
\caption{(a) Loschmidt echo decay in quantum systems with a chaotic classical limit. The crossover times are 
$t_c \sim \hbar / \eta$, delimiting the parabolic regime, and 
$t_s \sim \Gamma^{-1}\ln N$, associated with the transition into saturation. (b) Characteristic Decay Regimes of the Loschmidt Echo in Chaotic Systems. (c) Dependence of the Decay Rate $\Gamma$ on Perturbation Strength $\kappa$\label{fig:3paneles}}
\end{figure}
%%%%%%%%%%%%%%%%%%%%%%

\subsection{Theoretical Underpinnings}

The theoretical description of the Loschmidt echo benefits from multiple complementary approaches, each revealing different aspects of this complex phenomenon.
Its analysis relies on a powerful set of analytical and numerical techniques adapted from quantum chaos and many-body physics \cite{Jalabert2001, Gorin2006}. The primary analytical approaches are semiclassical analysis and Random Matrix Theory (RMT), each offering unique insights under different assumptions.

The semiclassical approach expresses quantum propagators as sums over classical trajectories, using the Van Vleck-Gutzwiller approximation \cite{Gutzwiller1990}.
When computing the Loschmidt echo, we must consider pairs of trajectories, those contributing to the forward evolution and their counterparts in the reversed evolution. The interference between these trajectory pairs determines the echo's behavior, with the accumulated action difference along nearby but distinct paths playing a particularly important role.
Applying this  \cite{Jalabert2001, Jacquod2001} leads to a complex expression involving a double sum over trajectory pairs from the forward and backward evolutions. To extract meaningful results, key approximations are employed: (i) the \emph{diagonal approximation}, which pairs identical trajectories (or those related by time-reversal) under the two Hamiltonians, justified by the shadowing theorem for small perturbations; and (ii) the classification of these pairs into correlated and uncorrelated families. The action difference for a paired trajectory is often given by $\Delta R_\gamma = -\kappa \int_0^t d\tau \, \Sigma(\mathbf{q}_\gamma(\tau))$ for a position-dependent perturbation \cite{Jalabert2001}. These steps reduce the problem to estimating phase accumulations and performing configurational averages, ultimately leading to decay expressions like the Lyapunov and Fermi-golden-rule regimes.

A significant advancement is the \emph{dephasing representation} \cite{Vanicek2004, Vanicek2006}, which bypasses the complex trajectory-search problem of standard semiclassics. It expresses the fidelity amplitude as
\[
m(t) = \int \mathrm{d}\mathbf{q} \, \mathrm{d}\mathbf{p} \, \exp\left(-\frac{i}{\hbar} \Delta R(\mathbf{q}, \mathbf{p}, t)\right)  W_0(\mathbf{q}, \mathbf{p}),
\]
where $W_0$ is the Wigner function of the initial state and $\Delta R$ is the accumulated action difference along the \emph{single} classical trajectory starting at $(\mathbf{q}, \mathbf{p})$ and $|m(t)|^2=M(t)$. This formulation treats the decay as an initial-value problem, greatly simplifying numerical calculations.

Random Matrix Theory provides a complementary, statistical framework \cite{Gorin2004, Stoeckmann2004}. It assumes the Hamiltonian (either $H_1$, the perturbation $\hsig$, or both) is a random matrix drawn from an appropriate ensemble (e.g., Gaussian Orthogonal Ensemble). This approach is justified for systems with ergodic classical dynamics, as it inherently incorporates universal spectral correlations. RMT successfully predicts the Fermi-golden-rule decay regime and has established formal links between the fidelity decay and the parametric cross-form factor of the spectrum \cite{Kohler2008}. However, its major limitation is the absence of finite classical time scales (e.g., the Lyapunov exponent $\lambda$ or the Ehrenfest time $t_E$), preventing it from describing regimes governed by these quantities.

Numerical simulations remain indispensable for testing analytical predictions and exploring regimes beyond current theory \cite{Cucchietti2002}. The primary challenge is the accurate computation of unitary time evolution under the Hamiltonians $H_1$ and $H_2$. Two highly efficient and stable algorithms are predominantly used: the \emph{Trotter-Suzuki decomposition} \cite{DeRaedt1996}, which approximates the propagator by a product of exponentials of simpler Hamiltonian terms, and the \emph{Chebyshev polynomial expansion} \cite{TalEzer1984}, which offers spectral accuracy by expanding the propagator in an orthogonal polynomial basis. These methods have been crucial in simulating the LE for systems ranging from quantum maps and billiards to spin chains.

\subsection{Characteristic Decay Regimes}

The decay of the LE is not universal but rather exhibits distinct dynamical regimes that depend critically on the interplay between the system's underlying classical dynamics, the initial state $|\psi_0\rangle$, and the strength and spatial structure of the perturbation $\Sigma$ defining $H_2 = H_1 + \kappa\Sigma$ \cite{Jalabert2001, Gorin2006}. The most profound understanding has emerged for systems that are classically chaotic, where the decay of the ensemble-averaged LE, $\overline{M(t)}$, follows a characteristic sequence of temporal  [See Fig. \ref{fig:3paneles} (a) and (b)]. An initial, universal \emph{parabolic decay} $\overline{M(t)} \simeq 1 - (\eta t/\hbar)^2$ is observed for all systems at very short times. This is followed by an intermediate-time asymptotic decay whose functional form is dictated by the perturbation strength. For weak perturbations, the decay follows a \emph{Fermi-golden-rule (FGR) regime}, $\overline{M(t)} \simeq \exp[-(\eta t/\hbar)^2]$, with a rate $\eta^2 \propto \kappa^2$ that is quadratic in the perturbation strength \cite{Jacquod2001}. The most striking regime emerges for stronger perturbations: a crossover to a perturbation-independent \emph{Lyapunov regime}, characterized by an exponential decay $\overline{M(t)} \simeq \exp(-\lambda t)$ where $\lambda$ is the classical Lyapunov exponent of the system \cite{Jalabert2001, Cucchietti2002} [see Fig. \ref{fig:3paneles} (c)]. This remarkable finding indicates that intrinsic chaos, rather than the external perturbation, becomes the dominant source of irreversibility. Ultimately, at times comparable to the Heisenberg time, the decay saturates at a value $\overline{M(t)} \sim N^{-1}$ governed by the effective size $N$ of the accessible Hilbert space \cite{gutierrez2009}.

This standard picture is richly modified by the nature of the perturbation. For \emph{global perturbations}, the Lyapunov regime persists until a breakdown strength $\kappa_b$ is reached. In contrast, for \emph{local perturbations} confined to a limited phase-space region, the exponential decay rate $\Gamma(\kappa)$ displays a more complex, non-monotonic dependence. It crosses over from the perturbative FGR regime ($\Gamma \sim \kappa^2$) to an \emph{escape-rate regime} ($\Gamma \to 2\gamma$) for very strong perturbations, where $\gamma$ is the classical escape rate into the perturbed region \cite{Goussev2008}. This transition is often accompanied by pronounced oscillations in $\Gamma(\kappa)$, the nature of which depends on the specific physical properties of the perturbation \cite{goussev2008loschmidt,ares2009loschmidt}.

The behavior is markedly different and less universal in systems with regular or mixed phase-space dynamics. Here, the LE decay is highly sensitive to the initial state's location relative to invariant classical structures such as Kolmogorov-Arnold-Moser (KAM) tori and islands of stability \cite{Prosen2002, Weinstein2005}. Commonly reported behavior includes a persistent \emph{Gaussian decay} for very weak perturbations and a faster \emph{algebraic decay} ($\overline{M(t)} \sim t^{-3d/2}$) for strong perturbations that appear random relative to the regular dynamics \cite{Prosen2002,jacquod2003anomalous}. Furthermore, one frequently encounters phenomena like temporary \emph{quantum freeze} and strong \emph{revivals} of the LE, which are direct manifestations of quantum coherence and the quasi-periodicity of the underlying motion \cite{prosen2003quantum}. This sensitivity makes a comprehensive classification exceptionally challenging for non-chaotic systems.

\subsection{Experimental Realizations of the Loschmidt Echo}

The experimental investigation of the Loschmidt echo has been successfully pursued across several physical systems, each offering unique advantages for studying different aspects of quantum time-reversal, insights and its limitations.

Nuclear magnetic resonance has served as the pioneering platform for Loschmidt echo experiments, beginning with Hahn's spin echo in the 1950s. NMR implementations have progressively increased in complexity, from reversing individual spin precession to addressing the challenging reversal of many-body dipolar interactions. The key developments include the magic echo technique by Rhim, Pines, and Waugh for reversing spin-spin interactions, and the polarization echo method developed by Ernst and collaborators, which enables the study of time-reversal for a locally injected spin excitation propagating through a many-spin network \cite{Hahn1950, Rhim1971, Levstein1998}. These experiments in ferrocene and cobaltocene crystals demonstrated either Gaussian or exponential decay of the polarization echo, with the crossover between these regimes providing evidence for environmentally induced decoherence mechanisms \cite{Usaj1998, Pastawski2000}.

Microwave cavities have provided an excellent experimental analogue for quantum billiards, leveraging the formal equivalence between the Helmholtz equation for electromagnetic waves and the Schr\"odinger equation. In these experiments, the scattering fidelity--obtained from the correlation of scattering matrix elements before and after a perturbation--has been shown to correspond to the Loschmidt echo for random initial states. This approach has enabled precise verification of theoretical predictions for both global and local Hamiltonian perturbations in chaotic systems, with measured decay rates showing excellent agreement with semiclassical calculations \cite{Schafer2005, Stockmann2005, Kober2011}.

The principles of LE have also been explored in classical wave systems, particularly through coda wave interferometry in elastic media. Experiments by Lobkis and Weaver \cite{Lobkis2003} demonstrated that 
temperature-induced perturbations in aluminum blocks cause measurable decorrelation of multiply scattered elastic waves. The resulting distortion of wave signals follows the same functional form as the fidelity decay predicted by random matrix theory, remarkably showing that this correspondence holds for both chaotic and regular systems \cite{Gorin2006}.

Ultracold atoms in optical potentials have emerged as a versatile platform for studying LE dynamics with exceptional control. Using atom interferometry techniques, researchers have measured quantities closely related to the LE by manipulating internal atomic states that experience different Hamiltonian evolution. These experiments have addressed challenges such as thermal decoherence through innovative pulse sequences and have explored fidelity behavior in chaotic billiards and kicked rotor systems \cite{Andersen2006, Wu2009, Ullah2011}. Recent work has even demonstrated the potential use of time-reversal protocols for cooling atomic matter waves \cite{Mart2008}.

While conceptually related, time-reversal mirrors in acoustics and electromagnetics represent a distinct approach to wave refocusing that has developed in parallel to quantum LE studies. These techniques exploit multiple scattering in complex media to achieve spatial and temporal focusing of classical waves, with applications ranging from medical imaging to telecommunications. The surprising effectiveness of single-channel time-reversal and the enhanced focusing in disordered media find natural explanation within a semiclassical framework that shares mathematical similarities with LE theory \cite{Fink1999, Calvo2008}.

\subsection{Applications and New Direccions}

The study of the Loschmidt echo extends far beyond academic interest, finding applications across multiple areas of modern physics. In quantum chaos research, it provides a crucial connection between classical and quantum descriptions of dynamical systems, quantifying how classical chaotic properties manifest in quantum behavior.
For quantum information science, the Loschmidt echo has become an essential tool for characterizing quantum operation robustness against imperfections and noise. The echo decay rate directly measures a protocol's sensitivity to perturbations -- critical information for designing fault-tolerant quantum algorithms.
Perhaps most profound are the applications in decoherence studies. By isolating the effects of specific perturbations, the Loschmidt echo helps distinguish between different decoherence sources and quantify their relative impacts. This capability proves particularly valuable for developing quantum technologies requiring prolonged coherence maintenance.

On a fundamental level, Loschmidt echo studies continue to illuminate the original reversibility paradox, helping us understand how macroscopic irreversibility emerges from microscopically reversible dynamics. Recent advances in understanding many-body echoes are opening new perspectives on this age-old problem.

In recent years, new directions in the research on the Loschmidt Echo have opened up.
Among the most promising one is the study of many-body systems with complex interactions. Understanding echo behavior in these systems could provide crucial insights into many-body localization and thermalization in isolated quantum systems.

The connection between LE behavior and thermalization mechanisms in isolated quantum systems has been extensively explored, revealing how the long-time dynamics and revival structure of the LE serve as sensitive probes of the Eigenstate Thermalization Hypothesis (ETH). In systems satisfying ETH, the LE typically decays to a small saturation value that encodes thermodynamic properties, while its temporal fluctuations reflect specific correlations between energy eigenstates \cite{Deutsch2018, DAlessio2016}.

A major breakthrough has been the establishment of the LE as a primary diagnostic tool for dynamical quantum phase transitions (DQPTs). Research has demonstrated that the rate function of the LE can exhibit non-analytic behavior in time, effectively serving as a dynamical order parameter that reveals universal scaling laws connected to equilibrium critical exponents, particularly in quench dynamics across quantum critical points \cite{Heyl2018, Calabrese2016}.

In the context of many-body localization, the LE has proven invaluable for distinguishing between thermalizing and localized phases. Unlike in thermalizing systems where the LE typically decays completely, in many-body localized systems it can saturate at a finite value, reflecting the characteristic preservation of local information and breakdown of thermalization that defines these phases \cite{Altman2018, Nandkishore2015}.

The relationship between LE decay and information scrambling has emerged as another rich area of investigation, with studies establishing concrete connections between the LE, out-of-time-order correlators (OTOCs), and quantum circuit complexity. The LE decay rate has been directly linked to the quantum butterfly effect, providing a quantitative measure of how rapidly local information becomes distributed throughout a quantum system \cite{Brown2016, Cotler2017}.

These theoretical advances have been complemented by significant experimental progress, with advanced quantum platforms including superconducting processors, trapped-ion systems, and ultracold atom setups enabling direct measurements of LE dynamics in regimes that are inaccessible to classical simulation. These experimental implementations have provided crucial validation of theoretical predictions concerning thermalization, many-body localization, and dynamical phase transitions \cite{Mi2022, Zhang2017}.

%%%%%%%%%%% OTOC  %%%%
\section{Out-of-time-ordered correlators: chaos and scrambling}
\label{sec:otoc}
{\subsection{Definition and basic properties}

Identifying dynamical signatures of chaos in quantum systems is fundamentally challenging. In classical mechanics, chaos is characterized by exponential sensitivity to initial conditions, quantified by Lyapunov exponents and directly visualized through the divergence of nearby trajectories. In quantum mechanics, however, the notion of a trajectory is absent, and the unitary nature of time evolution preserves Hilbert-space distances between states. As a consequence, classical definitions of instability cannot be straightforwardly translated to the quantum domain.

Early approaches to quantum chaos therefore focused on \emph{spectral statistics}, which successfully distinguish integrable from chaotic systems through level correlations. While extremely powerful, these indicators are inherently static and do not directly probe dynamical processes such as information spreading or relaxation. This limitation has motivated the development of dynamical diagnostics capable of capturing how quantum systems process and redistribute information over time.

Out-of-time-ordered correlators (OTOCs) have emerged as one of the most prominent such diagnostics. Their central idea is to quantify how the time evolution of an initially simple operator affects its commutation relations with another operator at later times. In this way, OTOCs provide a notion of dynamical sensitivity that is fully compatible with quantum mechanics while retaining a clear connection to classical instability in appropriate limits.

In contrast to the Loschmidt echo, which probes the stability of \emph{states} under imperfect time reversal, OTOCs probe the instability of \emph{operators} under forward unitary evolution. They therefore shift the emphasis from reversibility to scrambling: rather than asking how well an initial state can be reconstructed, they quantify how rapidly initially localized information becomes inaccessible to local measurements. 

These two perspectives --echo dynamics and operator growth-- are in fact closely related. It has been shown that, under suitable averaging procedures and appropriate choices of operators, OTOCs can be expressed as thermal averages of Loschmidt echo signals \cite{yan_zurek2020}. This correspondence provides a direct conceptual bridge between reversibility-based diagnostics and measures of scrambling.

A commonly used definition of the OTOC is based on the averaged squared commutator (SC),
\begin{equation}
\label{otocdef1}
C_{VW}(t) = \langle [\hW(t),\hV]^\dagger [\hW(t),\hV] \rangle ,
\end{equation}
where $\hW(t)=e^{iHt/\hbar} \hW e^{-iHt/\hbar}$ is the Heisenberg-evolved operator. The expectation value is typically taken with respect to a thermal density matrix $\rho$, so that $\langle \cdot \rangle = \mathrm{tr}(\rho \cdot)$. In many situations, especially in theoretical studies, one considers the infinite-temperature limit $\rho = \mathbb{I}/N$, or averages over random pure states.

The physical meaning of Eq.~\eqref{otocdef1} becomes transparent when considering two initially commuting operators, $[\hW,\hV]=0$. As time evolves, $\hW(t)$ spreads over the operator algebra, and its support may eventually overlap with that of $\hV$. The growth of the commutator therefore directly measures how operator evolution generates nonlocal correlations—a hallmark of information scrambling.

Expanding Eq.~\eqref{otocdef1}, one obtains
\begin{equation}
{C}_{\hV\hW}(t) = {D}_{\hV\hW}(t) + {I}_{\hV\hW}(t) - 2 \,  \mathrm{Re}\{{F}_{\hV\hW}(t)\},
\end{equation}
where $D$ and $I$ are time-ordered correlators, while
\begin{equation}
{F}_{\hV\hW}(t) = \left \langle \hW_t^{\dagger} \hV^{\dagger} \hW_t \hV \right \rangle
\end{equation}
is the genuine out-of-time-ordered correlator. The nontrivial time ordering of $F(t)$--in which operators at different times are interleaved-- is precisely what makes OTOCs sensitive to operator growth and scrambling dynamics.

For Hermitian unitary operators, the expression simplifies considerably to
\begin{equation}
C_{\hV\hW}(t) = 2\left(1 - \mathrm{Re}\{F_{\hV\hW}(t)\}\right),
\end{equation}
a form that is particularly convenient in both analytical calculations and experimental implementations.

In practical applications, the choice of operators $\hV$ and $\hW$ plays a crucial role. Typically, one selects local operators in order to probe the spreading of initially localized perturbations. This allows one to track the emergence of effective light-cone structures and to define quantities such as the butterfly velocity.

At finite temperature, it is common to introduce a \emph{regularized} version of the OTOC by inserting $\rho^{1/2}$ symmetrically between operators. While this modification is often assumed not to affect the asymptotic long-time behavior, it can lead to quantitative differences at intermediate times and may influence the extraction of growth rates \cite{liao2018nonlinear,romero2019,sahu2020}. The choice between regularized and unregularized definitions therefore depends on the specific physical context.

Overall, OTOCs provide a flexible and powerful framework for probing operator growth, information scrambling, and dynamical instability in quantum systems, forming a natural counterpart to both Loschmidt echo diagnostics and classical measures of chaos.}
%%%%%%%%%%%%%%%%%%%%%%
\begin{figure}
\begin{center}
\includegraphics[width=0.5\linewidth]{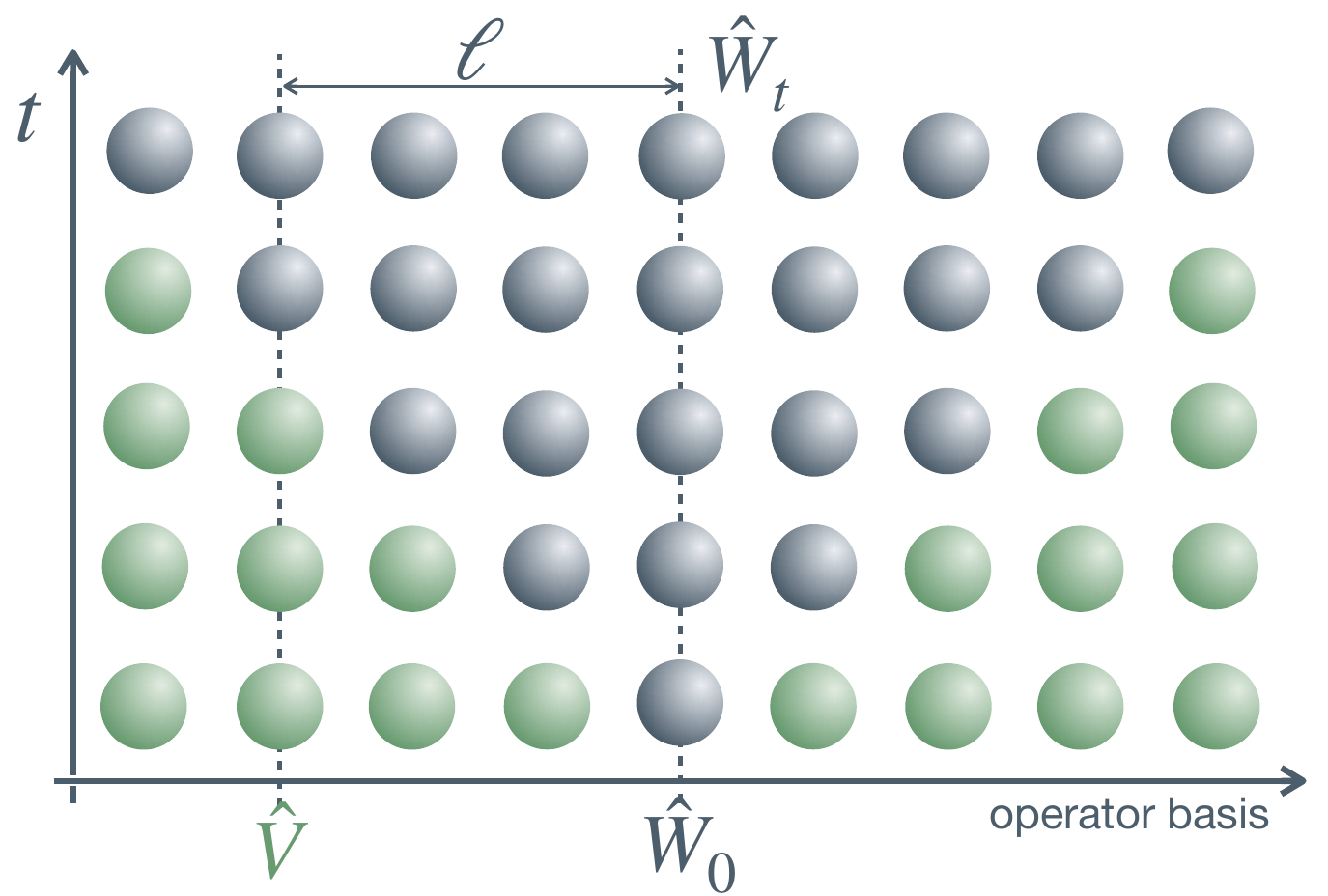}
\end{center}
\caption{Sketch illustrating the spreading of an initially localized operator $\hat{W}$, acting on the central site of a one-dimensional system (e.g., a spin chain), under Heisenberg evolution $\hat{W}(t)$. A second operator $\hat{V}$, located at a distance $l$ from the center, is initially unaffected, but becomes influenced once it enters the propagation cone of $\hat{W}(t)$ (gray shaded circles).\label{fig:1otoc}}
\end{figure}
%%%%%%%%%%%%%%%%%%%%%%%%%%%%

\subsection{Physical interpretation and operator growth}
The growth of the squared commutator is associated with information scrambling\cite{shenker2015stringy}, caused by spreading of quantum information across operator degrees of freedom. A usual depiction is shown in Fig.~\ref{fig:1otoc}. A one-dimensional system, two spacelike-separated, local operators $\hat{W}$ and $\hat{V}$ (i.e. $[\hat{W},\hat{V}]$=0, initially), with $\hat{W}$ acting in the middle of the chain, and $\hat{V}$ at distance $\ell$. The time evolution of $\hat{W}$ by Hamiltonian $\hat{H}$ can be written in Baker-Campbell-Haussdorff form as
\begin{equation}
\label{eq:opevol}
\hat{W}_t=\sum_{k=0}^{\infty} \frac{(i t)^k}{k!} \underbrace{[\hat{H}, \ldots[\hat{H}}_k, \hat{W}] \ldots].
\end{equation}
This equation allows to infer that as time grows, the presence of n-particle interactions produces a growth in the number of sites involved. In Fig.~\ref{fig:1otoc}, this growth is seen in ``light cone'' fashion. Initially one site is affected by $\hat{W}$, and as evolution in time takes place according to $H$, more terms get involved, until at some point the site associated with $\hat{V}$ is reached. The nonvanishing commutator probes this effect.

Fast scrambling depends exponentially in the variable $t-\ell/v_b$, $C(t)$ is very small. The butterfly velocity $v_b$ determines the slope of the light cone and is limited by the Lieb-Robinson bound \cite{LiebRobinson72,roberts_swingle_2016}. The dependence of the commutator growth can of course be other than exponential \cite{Xu2019Butterfly,Xu2019Locality,swingle2020}

\subsection{Relation to classical chaos}
\label{relation}
The OTOC can, under certain conditions, be tightly related to the Loschmidt echo \cite{kurchan2018quantum,schmitt2018,schmitt2019,Sanchez2021,yan_zurek2020}. It is, thus, not surprising to find tight relations in the semiclassical limit to classical characteristics of chaos.
In systems admitting a semiclassical limit, the exponentioal growth of the SC can be directly related to classical measures of instability. In this regime, commutators can be replaced by Poisson brackets,
\begin{equation}
\frac{1}{i\hbar}[W(t),V] \rightarrow \{W(t),V\},
\end{equation}
so that the OTOC probes the sensitivity of classical observables to initial conditions.

If we choose $\hW=\hat{X}$ and $\hV=\hat{P}_x$, the classical limit of the commutator $[\hat{X},\hat{P}_x]$ is
\begin{equation}
\mathrm{i} \hbar\left\{X(t), P_X(0)\right\}=i \hbar \frac{\partial X(t)}{\partial X(0)}
\end{equation}
In the fully chaotic case, the exponential sensitivity to initial conditions implies that
\begin{equation}
\partial X(t) / \partial X(0) \sim \exp (\lambda t),
\end{equation}
where
$\lambda$ is the largest classical Lyapunov exponent. 
Therefore, in the quasi-classical limit the SC, for small times, behaves as
\begin{equation}
\label{explambdat}
C_{\hat{X},\hat{P}_x}(t) \sim \hbar^2 e^{2\lambda t}.
\end{equation}
This regime persists only up to a finite time scale, typically identified with the Ehrenfest or scrambling time $t^*$, beyond which quantum interference effects suppress further exponential growth \cite{rammensee2018}.
{This semiclassical correspondence highlights a conceptual bridge with the Loschmidt echo discussed earlier. While the LE translates classical instability into decay of state overlap under imperfect reversal, the OTOC translates it into growth of noncommutativity under forward dynamics. Both quantities therefore encode sensitivity to perturbations, but at complementary levels: the LE at the level of states, and the OTOC at the level of operators.}

%%%%%%%%%%%%%%%%%%%%%%
\begin{figure}
\begin{center}
\includegraphics[width=0.85\linewidth]{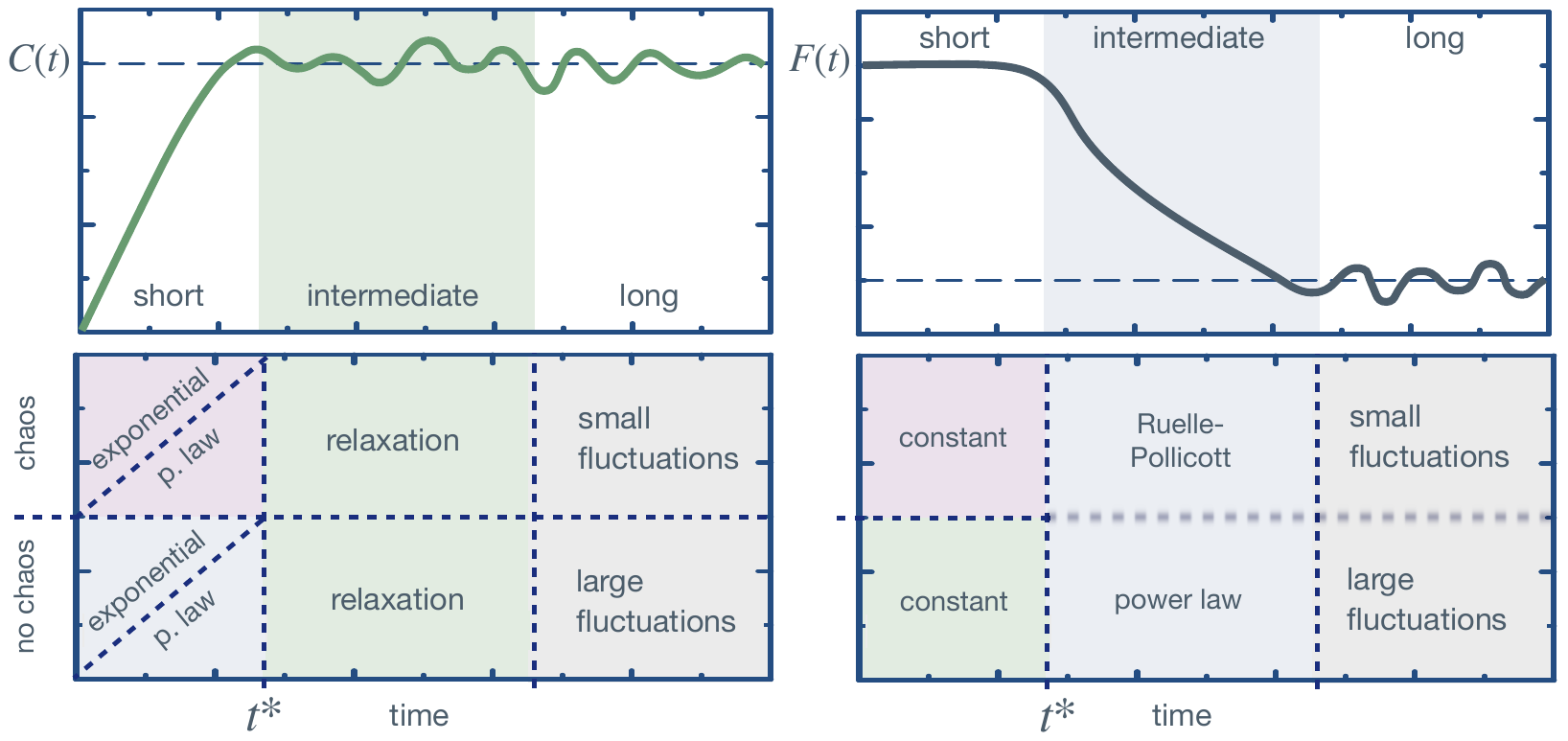}
\end{center}
\caption{Schematic depiction of the time regimes of $C(t)$ and $F(T)$. The SC $C(T)$ grows initially, when scrambling occurs up to scrambling time $t^*$. After that $C(t)$ relaxes at a rate determined mainly by the decay of $F(t)$ (right panel). After the $F(t)$ saturates to a value that depends on the dynamics and system size. In both cases, the study of fluctuations after saturation give information about the dynamics \cite{FortesPRE2019}.  Contrary to common lore, the initial exponential growth of the SC is not a clear signature of chaos, nor is power law growth a signature of absence of chaos (left panel). Exponential decay of the OTOC (Ruelle-Pollicott regime) is a more reliable signature of quantum chaos (right panel).  The blurry-dashed line in the right panel indicates a crossover rather than a sharp separation.  \label{fig:regimes}}
\end{figure}
%%%%%%%%%%%%%%%%%%%%%%%%

\subsection{Time regimes of OTOC dynamics}

The dynamics of the squared commutator (SC) and the corresponding OTOC can be usefully organized into three generic time regimes: an initial growth regime, an intermediate relaxation regime, and a long-time saturation regime. This division is schematic (see Fig.~\ref{fig:regimes}) and depends on system size and dynamical properties, but it provides a useful framework for interpreting operator growth.

At very short times, the OTOC exhibits perturbative behavior determined by the operator algebra and the local structure of the Hamiltonian. This regime is generally non-universal and depends sensitively on the chosen operators and initial state. We do not focus on this initial perturbative window.

The first physically significant regime corresponds to the growth of the commutator, reflecting the spreading of $\hW(t)$ over the operator basis as dictated by Eq.~(\ref{eq:opevol}). This spreading encodes quantum information scrambling. The SC increases until a characteristic time $t^*$, often called the scrambling time, after which it ceases its growth and begins to fluctuate around a saturation value. In systems with a well-defined semiclassical limit, $t^*$ typically coincides with the Ehrenfest time $t_E$, the timescale over which a localized wave packet explores the available phase space. In a Hilbert space of dimension $N$, one expects $t_E \sim \frac{\ln N}{\lambda}$, where $\lambda$ is the classical Lyapunov exponent \cite{DimaScholar_Ehrenfest}.

The renewed interest in this growth regime was largely driven by studies of quantum chaos in black-hole physics \cite{shenker2014black,polchinski2015,Cotler2017,jahnke2019chaos}. In so-called fast scramblers \cite{sekino2008fast}, the SC exhibits exponential growth,
\begin{equation}
\langle \hW_t,\hV \rangle \sim e^{\Lambda t},
\end{equation}
with $\Lambda$ identified as a quantum Lyapunov exponent \cite{maldacena2016,hosur2016chaos}. In systems admitting a semiclassical limit, agreement between classical and quantum Lyapunov exponents has been verified in several contexts, including strongly chaotic one-body systems \cite{JGMW2018,Rozenbaum2017,OTOC_gato_PRL,lakshminarayan2019out,morita2021extracting}, the Dicke model \cite{chavez2019quantum}, and the inverted harmonic oscillator \cite{ali2020chaos,morita2021extracting}. This early-time exponential growth has been widely interpreted as the quantum analogue of the butterfly effect \cite{Lorenz1963,Lorenz1972}, motivating the notion of a butterfly velocity governing spatial spreading \cite{Roberts2015,roberts_swingle_2016,Swingle2017}.

However, the association between exponential SC growth and quantum chaos is not universal. One-dimensional spin systems with short-range interactions often exhibit power-law growth of the SC \cite{craps2020lyapunov,Motrunich2018,Riddel2019,FortesPRE2019}, even in parameter regimes where spectral statistics follow random matrix theory \cite{bohigas}. Conversely, exponential growth can arise in integrable systems possessing unstable fixed points or saddle structures in phase space. In such cases, local exponential separation of trajectories near unstable manifolds can produce rapid scrambling without global chaos \cite{Fazio2018,XuPRL2020,PilatowskiPRE2020,HummelPRL2019,Bhattacharyya2022,ali2020chaos,hashimoto2020exponential,Bhattacharyya2021multi,kidd2021saddle}. These observations underscore that early-time exponential growth alone does not uniquely diagnose chaos.

Beyond the scrambling time, the dynamics enters an intermediate regime characterized by relaxation. After the Lyapunov-driven growth terminates, the SC fluctuates around its eventual plateau while the OTOC begins to decay. In integrable systems, this decay is typically algebraic. In mixed systems, a crossover from exponential to power-law decay can occur \cite{notenson2023classical}. In strongly chaotic systems, the decay is exponential and reflects relaxation toward equilibrium.

This regime can be understood in terms of the distinction between stretching and mixing. While Lyapunov exponents characterize short-time instability (stretching), mixing governs long-time relaxation. In compact phase spaces, mixing arises when stretched trajectories fold repeatedly onto themselves, as in the baker's map or Smale's horseshoe. In classical systems, decay of correlations is governed by the spectrum of the Perron-Frobenius operator, whose complex eigenvalues --the Ruelle-Pollicott resonances (RPRs)-- determine relaxation rates \cite{pollicott1985rate,ruelle1986,ruelle1987resonances}. The real parts of the RPRs control decay rates, while the imaginary parts encode oscillatory components.

Because the OTOC is itself a correlation function, it naturally reflects this mixing regime. In quantum systems with a clear classical counterpart, the intermediate-time decay toward saturation is governed by the leading RPR, as explicitly demonstrated for the perturbed cat map \cite{OTOC_gato_PRL}. More generally, this separation between an early Lyapunov regime and a later RPR-dominated regime has been established in quantum chaotic maps and semiclassical systems \cite{OTOC_gato_PRL,FortesPRE2019}.

For many-body systems lacking a classical limit, identifying quantum analogues of RPRs remains an active area of research. Early work based on operator-space coarse graining showed that eigenvalues of appropriately defined propagators can be interpreted as quantum RPRs \cite{Prosen2002Ruelle,Prosen2004Ruelle,Prosen2007Chaos}. More recent approaches exploit translational symmetry and quasimomentum decomposition to resolve resonances within symmetry sectors \cite{znidaric2024momentum}. An alternative perspective is provided by the Liouvillian gap,
\begin{equation}
g = - \max_{\alpha \neq 0} \operatorname{Re} \lambda_\alpha ,
\end{equation}
where $\lambda_\alpha$ are the nonzero complex eigenvalues of the Liouvillian superoperator. In the limits of infinite system size and vanishing dissipation, the Liouvillian gap coincides with the logarithm of the largest quantum RPR in solvable models \cite{Jacoby2025,Zhang2024RPR,Yoshimura2025}. The correspondence between the OTOC decay rate and the Liouvillian gap was explicitly demonstrated for the kicked Ising spin chain in Ref.~\cite{Duarte2026Ruelle}.

At sufficiently long times, the SC saturates due to the finite Hilbert-space dimension $D$. In chaotic systems, the late-time plateau is well described by Haar-random estimates. For infinite-temperature averages and traceless operators normalized as $\tr(\hat V^2)=\tr(\hat W^2)=D$, one finds
\begin{equation}
\overline{F(t)} \to 0, 
\qquad
\overline{C(t)} \to 2\left(1-\frac{1}{D}\right),
\end{equation}
up to corrections of order $1/D$ \cite{RobertsYoshida2017}. From the viewpoint of the eigenstate thermalization hypothesis, this saturation reflects effective ergodicity in operator space and the loss of memory of the initial perturbation.

Importantly, the long-time regime is not trivial. The magnitude and structure of temporal fluctuations around the plateau carry diagnostic information. In chaotic systems, fluctuations are strongly suppressed and consistent with random-matrix predictions. In integrable or mixed systems, fluctuations are enhanced and more structured, reflecting incomplete scrambling. This distinction was quantified in Ref.~\cite{FortesPRE2019}, where long-time OTOC fluctuations were shown to serve as a dynamical probe of chaos. In many-body localized phases, deviations from Haar scaling and persistent fluctuations further reflect the absence of thermalization.

Taken together, these regimes show that OTOCs encode within a single observable the key dynamical ingredients of quantum chaos: early-time instability, intermediate-time mixing, and long-time ergodic or nonergodic structure.
{From this perspective, OTOCs provide a temporal narrative closely paralleling that of the Loschmidt echo: an initial regime dominated by local instability (analogous to Lyapunov growth), an intermediate regime governed by relaxation and mixing, and a long-time regime where saturation and fluctuations reflect the effective size and structure of the accessible Hilbert space. This unified organization helps connect scrambling, mixing, and ergodicity within a single operational probe.}

\subsection{Many-body systems: integrability and localization}

Out-of-time-ordered correlators (OTOCs) provide a dynamical probe of operator spreading and information scrambling in interacting quantum systems. 
Beyond semiclassical settings with well-defined Lyapunov exponents, their behavior in genuinely many-body systems where no simple classical analog exists reveals clear distinctions between thermalizing, integrable, and many-body localized (MBL) regimes \cite{otocScholarpedia}.

In chaotic many-body systems, local perturbations spread ballistically, generating light-cone-like structures and leading to efficient scrambling. 
The OTOC approaches its saturation value smoothly, and long-time fluctuations are strongly suppressed, consistent with ergodic exploration of Hilbert space. 
Integrable systems behave qualitatively differently. 
The presence of an extensive set of conserved quantities constrains operator spreading, leading to polynomial growth or persistent oscillations rather than sustained exponential behavior. 
Even at long times, fluctuations remain significant, reflecting structured dynamics and the absence of random-matrix-type spectral correlations \cite{FortesPRE2019}. 
Thus, while spectral statistics distinguish integrable and chaotic systems in the energy domain, OTOCs reveal their dynamical differences directly through operator growth.

Many-body localization represents a more extreme departure from thermal behavior. 
In disordered interacting systems, transport is suppressed, entanglement entropy grows only logarithmically in time, and local memory persists indefinitely \cite{Kim2014,chen2016universal,smith2019logarithmic}. 
Correspondingly, operator spreading becomes logarithmically slow: the effective light cone scales as $\log t$ rather than linearly in time \cite{fan2017out}. 
OTOCs in the MBL phase therefore exhibit slow, often power-law decay and enhanced long-time structure. 
Unlike conventional two-point correlators, OTOCs directly capture the interaction-induced dephasing that drives this slow propagation \cite{Huang2016}.

Across the transition from ergodic to localized behavior, OTOCs interpolate between ballistic scrambling and logarithmic spreading \cite{chen2017out}. 
They therefore provide a unified dynamical diagnostic: rapid scrambling in chaotic systems, constrained quasiparticle propagation in integrable systems, and slow dephasing in MBL phases.

Taken together, these results emphasize that operator growth is not synonymous with chaos. 
OTOCs probe not only exponential sensitivity but also ergodicity, conservation laws, and localization, offering a coherent dynamical framework for understanding many-body quantum phases.

\subsection{OTOCs in experiments}

The experimental measurement of out-of-time-ordered correlators is intrinsically challenging due to their nonstandard time ordering, which typically requires either controlled reversal of the system dynamics or the use of auxiliary degrees of freedom. Over the past decade, however, several complementary strategies have been developed and implemented across different quantum platforms. The following discussion highlights representative examples, without attempting an exhaustive survey of the rapidly expanding literature. Together, these experiments illustrate how OTOCs have evolved from theoretical diagnostics to experimentally accessible probes of many-body dynamics.

Trapped-ion systems were among the first platforms to access OTOCs experimentally. In a landmark experiment, Ref.~\cite{Garttner2017} implemented a protocol based on reversing many-body dynamics in a long-range Ising spin simulator containing more than 100 ions in a Penning trap. By measuring a family of OTOCs, the buildup of multi-body correlations was directly observed, providing quantitative insight into scrambling and operator growth. Subsequent work extended these ideas in both analog and digital directions. Ref.~\cite{joshi2020quantum} demonstrated scrambling in a 10-qubit trapped-ion simulator using randomized measurement protocols that bypass explicit time reversal, while Ref.~\cite{green2022experimental} measured finite-temperature OTOCs on a digital trapped-ion processor via thermofield double state preparation, probing the temperature dependence of scrambling rates. These developments established trapped ions as versatile systems for exploring operator spreading across different regimes.

Nuclear magnetic resonance (NMR) experiments provided some of the earliest demonstrations of OTOC measurements in controllable quantum systems. Ref.~\cite{Li2017} reported measurements of OTOCs in an Ising spin chain, distinguishing integrable from nonintegrable behavior and extracting a butterfly velocity associated with correlation propagation. The flexibility of NMR echo protocols has also enabled the study of localization phenomena. Ref.~\cite{wei2018exploring} employed time-reversal techniques to differentiate many-body localization from Anderson localization, extracting correlation lengths from out-of-time-ordered signals. Further work showed that OTOCs can serve as probes of both equilibrium and dynamical phase transitions \cite{Nie2020experimental}, and that nonlocal variants can quantify the sensitivity of system-environment correlations to perturbations \cite{MohamadPRR2020}. More recently, global OTOCs measured in solid-state nuclear-spin systems were shown to agree quantitatively with stochastic theoretical models \cite{zhou2023operator}. These experiments underscore the close connection between OTOCs, echo protocols, and controlled reversibility.

Superconducting circuits have enabled OTOC measurements in programmable architectures with increasing size and flexibility. Ref.~\cite{braumuller2022probing} implemented OTOCs in a $3\times3$ Bose-Hubbard lattice realized on a superconducting processor, combining digital-analog simulation with coherent time reversal to study information propagation and Loschmidt echoes. Floquet-engineered superconducting chains have further revealed light-cone-like operator spreading and scrambling behavior even near integrable regimes \cite{Zhao2022probing}. More recently, higher-order OTOCs have been measured directly on superconducting quantum processors using randomized Pauli insertions during evolution \cite{google2025Observation}, demonstrating access to multi-point correlations beyond the standard four-point function.

Taken together, these experimental advances show that OTOCs can be accessed through time-reversal schemes, interferometric protocols, or randomized measurements, depending on the platform. Across trapped ions, NMR systems, and superconducting circuits, OTOCs have provided direct experimental windows into scrambling, operator spreading, localization, and dynamical phase transitions. They have thus become practical and versatile tools for characterizing quantum chaos and many-body dynamics in controllable quantum devices.

\subsection{New directions: Higher-order OTOCs}
The standard OTOC probes the second moment of operator growth through four-point correlators such as 
\begin{equation}
F(t)=\langle W(t) V W(t) V \rangle .
\end{equation}
More generally, one may consider higher-order out-of-time-ordered correlators of the form
\begin{equation}
F_n(t)=\langle (W(t)V)^n \rangle ,
\end{equation}
or related multi-point generalizations. 
These quantities probe higher moments of the operator-growth distribution and therefore encode information beyond the average scrambling captured by the standard OTOC.

While the conventional four-point OTOC measures the onset of instability and the spreading of operator support, higher-order correlators are sensitive to the full statistical structure of operator evolution. 
In chaotic systems, where dynamics effectively randomizes operators in Hilbert space, one expects higher-order OTOCs to approach Haar-random values at long times. 
Deviations from these expectations can reveal residual structure due to conservation laws, integrability, or localization.

Recent theoretical work has emphasized that higher-point OTOCs provide access to refined aspects of scrambling, including the distribution of operator weights and multi-scale relaxation processes \cite{Pappalardi2024}. 
In particular, they can distinguish between simple exponential growth and more intricate operator dynamics that are invisible at the level of the standard four-point function.

Experimentally, higher-order OTOCs are beginning to become accessible. 
In a recent superconducting quantum processor experiment, second-order OTOCs were directly measured, revealing interference effects that persist even after conventional OTOCs have saturated \cite{GoogleOTOC2025}. 
These results demonstrate that higher-order correlators retain sensitivity to coherent many-body structures at long times and can probe dynamical features inaccessible to lower-order diagnostics.

Thus, higher-order OTOCs extend the scrambling paradigm from average growth to the full hierarchy of dynamical correlations, offering a richer characterization of quantum chaos and many-body complexity.

%%%%%%%%%%%%%Krylov
\section{Exploring Quantum Complexity through the Krylov Approach}

The discussion above has emphasized operator growth and information scrambling as central dynamical features of quantum chaos, as captured by OTOCs. A complementary perspective consists in describing this growth not through correlation functions, but in terms of how a state or operator spreads in a dynamically generated basis.

A novel approach to measuring the complexity of quantum evolutions has emerged in recent years using \textbf{Krylov subspaces} and \textbf{Krylov complexity (K-complexity)} \cite{Parker,nandy2025quantum,rabinovici2025krylov}. Within this framework, the time evolution is expressed in a basis constructed recursively from the Hamiltonian (or Liouvillian), allowing one to interpret operator growth as spreading in Krylov space.
%A novel approach to measuring the complexity of quantum evolutions has emerged in recent years using \textbf{Krylov subspaces} and \textbf{Krylov complexity (K-complexity)} \cite{Parker,nandy2025quantum,rabinovici2025krylov}.

The Krylov method is a numerical technique initially developed to efficiently approximate the exponentiation of matrices \cite{hochbruck1997krylov,parlett1998symmetric}. This is particularly useful for large systems, as commonly encountered in quantum mechanics. It has been adapted to study the complexity of quantum evolution for both states and operators \cite{Parker,Rabinovici2021,Balasubramanian2022}.

The core idea involves mapping the system's evolution onto the dynamics of a single-particle wave function in a constructed space called the \textbf{Krylov space}. This mapping results in an effective one-dimensional tight-binding model, where the hopping coefficients are given by the so-called \textbf{Lanczos coefficients} ($a_n$ and $b_n$) \cite{rabinovici2025krylov}. The Krylov basis is argued to be particularly effective at minimizing the spread of the quantum evolution \cite{Rabinovici2022,Balasubramanian2022}.

\textbf{K-complexity}, denoted as $C_K$ [16, 18] or KC [9, 19], is defined within this framework as the average position of the one-particle wave function along the Krylov basis [16, 18, 20]. 
This measure quantifies how far a state or operator spreads over a specially constructed basis—the \textit{Krylov basis}—which is designed to capture the growth of complexity under time evolution.
Equivalently, this complexity measure can be understood as the average dimension of the Krylov subspace required to represent the time evolution of the initial state [18, 21].

Moreover, a ``universal operator growth hypothesis'' (see Ref.~\cite{Parker}) posits that, under chaotic Hamiltonian dynamics, Krylov complexity grows exponentially, with a rate that upper-bounds those of other operator-growth measures, such as the OTOC. The latter, in turn, reflects aspects of the underlying classical dynamics (when a classical limit exists), since its growth rate can be related to the Lyapunov exponent.
%%%%%%%%%%%%%%
\subsection{The Lanczos Algorithm and Krylov Basis Construction for States and Operators}

The foundation of Krylov complexity lies in the \textit{Lanczos algorithm}, which constructs an orthonormal basis for the Krylov subspace and generates the \textit{Lanczos coefficients} that dictate the dynamics in this basis.

Given an initial state $|\psi_0\rangle$ and a time-independent Hamiltonian $H$, the Krylov subspace for state evolution is defined as
\[
\mathcal{K} = \text{span}\{|\psi_0\rangle, H|\psi_0\rangle, H^2|\psi_0\rangle, \dots, H^{K-1}|\psi_0\rangle\},
\]
where $K$ is the dimension of the subspace (typically equal to the Hilbert space dimension $D$ when no degeneracies are present).

The Lanczos algorithm proceeds recursively, starting with $|K_0\rangle = |\psi_0\rangle$, $b_0 = 0$, and $|K_{-1}\rangle = 0$. The Krylov basis elements are defined through
\begin{equation}
|A_n\rangle = (H - a_{n-1})|K_{n-1}\rangle - b_{n-1}|K_{n-2}\rangle,\quad  |K_n\rangle = \frac{|A_n\rangle}{b_n},
\end{equation}
where
\begin{equation}
a_n = \langle K_n|H|K_n\rangle,\quad
b_n = \sqrt{\langle A_n|A_n\rangle}.
\end{equation}
The algorithm terminates when $b_n = 0$ (which occurs when $n = K$). The coefficients $\{a_n\}$ (diagonal) and $\{b_n\}$ (off-diagonal) are known as the \textit{Lanczos coefficients}. In the Krylov basis, the Hamiltonian takes a tridiagonal form,
\[
H|K_n\rangle = a_n|K_n\rangle + b_{n+1}|K_{n+1}\rangle + b_n|K_{n-1}\rangle.
\]

The time-evolved state can be expanded as
\[
|\psi(t)\rangle = \sum_{n=0}^{K-1} \psi_n(t) |K_n\rangle,
\]
where the amplitudes $\psi_n(t)$ satisfy a discrete Schr\"odinger equation,
\[
i\partial_t \psi_n(t) = a_n \psi_n(t) + b_n \psi_{n-1}(t) + b_{n+1} \psi_{n+1}(t).
\]

An analogous construction applies to operator dynamics in the Heisenberg picture, where one uses the Liouvillian superoperator $\mathcal{L} = [H, \cdot]$. The corresponding Krylov subspace is
\[
\mathcal{K}_{\text{op}} = \text{span}\{\mathcal{O}, \mathcal{L}\mathcal{O}, \mathcal{L}^2\mathcal{O}, \dots\}.
\]

The Lanczos algorithm for operators follows the same recursive structure, yielding orthonormal operator basis elements $\{|\mathcal{O}_n\rangle\}$ and Lanczos coefficients $\{b_n\}$ (with $a_n = 0$ for Hermitian operators). The corresponding amplitudes $\phi_n(t)$ satisfy
\[
\partial_t \phi_n(t) = b_n \phi_{n-1}(t) - b_{n+1} \phi_{n+1}(t).
\]
%%%%%%%%%%%%%%%%%
%%
\subsection{Krylov Complexity and its connection with quantum chaos}

The evolution of states and operators in the Krylov basis makes it possible to define a quantity referred to as Krylov complexity, which corresponds to the expectation value of the position operator in Krylov space.  This quantity is defined for states as,
\begin{equation}
    C_{\mathcal{K}}(t) = \sum_{n=0}^{K-1} n |\psi_n(t)|^2
\end{equation}
and for operators,     
\begin{equation}
        \mathcal{K}_{\mathcal{C}}(t) = \sum_{n=0}^{K-1} n |\phi_n(t)|^2
\end{equation}

The long-time saturation value is given by:
\[
\overline{C_{\mathcal{K}}} = \sum_{n=0}^{K-1} n Q_{0n}, \quad \text{where} \quad Q_{0n} = \sum_i |\langle e_i|\psi_0\rangle|^2 |\langle K_n|e_i\rangle|^2,
\]
and $\{|e_i\rangle\}$ are the energy eigenstates.

Krylov complexity has emerged as a fundamental measure of operator growth in quantum systems, providing deep insights into quantum chaotic dynamics. Since Parker's seminal work \cite{Parker} introducing Krylov complexity, extensive research has revealed profound connections between this quantity and quantum chaotic behavior. The most robust and universal signatures emerge in both the early-time growth and long-time saturation behavior of complexity, offering a complementary perspective to traditional measures like spectral statistics and out-of-time-order correlators \cite{maldacena2016}.

\begin{figure}
\begin{center}
\includegraphics[width=0.9\linewidth]{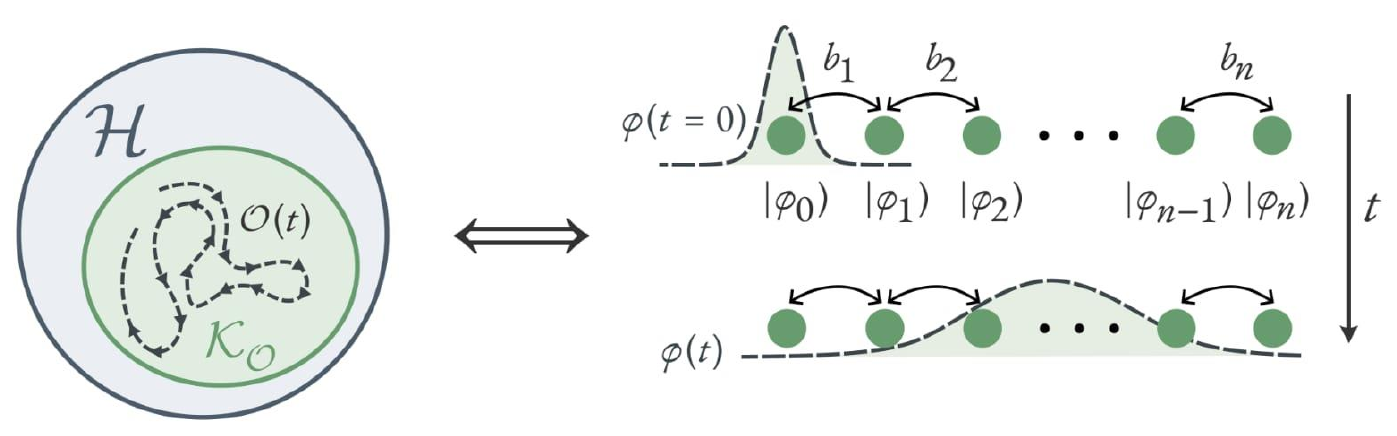}
\end{center}
\caption{
Time evolution of an operator $O(t)$ from the perspective of the Lanczos basis. In Liouville space it can be viewed as the diffusion of an initial state $ \lvert 0 $, representing the operator \( O \), fully localized at the leftmost site of a virtual tight-binding chain in operator space. In this picture, the tridiagonal Lanczos matrix \( T \) approximates the Liouvillian \( \mathcal{L} \), where the off-diagonal elements \( \beta_i \) act as hopping amplitudes between neighboring sites and the diagonal elements \( \alpha_i \) correspond to local on-site potentials (not shown). Truncating the Lanczos basis to \( N \) vectors effectively cuts the chain at site \( N \) .
\label{fig1:krylov}}
\end{figure}

In quantum chaotic systems, the Lanczos coefficients $b_n$ exhibit a characteristic tripartite structure: initial linear growth $b_n \sim \alpha n$, followed by a distinct plateau region, and eventual decline \cite{Rabinovici2021,Rabinovici2022}. This structured progression directly enables exponential growth in Krylov complexity during early times $\mathcal{K}(t) \sim e^{\lambda_K t}$, subsequently transitioning to linear growth before final saturation. The growth rate $\lambda_K$, known as the Krylov exponent, has been shown to saturate the fundamental bound $\lambda_K \leq 2\pi T$ at finite temperature \cite{Parker}, mirroring the behavior of Lyapunov exponents in chaotic systems.

The underlying mechanism connecting these patterns to chaos stems from spectral statistics and their manifestation in the Krylov space. Quantum chaotic systems, characterized by Wigner-Dyson statistics and strong level repulsion \cite{berry1977}, generate Lanczos coefficients that evolve smoothly and follow predictable scaling laws \cite{dymarsky2021}. This regularity facilitates extensive exploration of the Krylov space, allowing complexity to reach parametrically larger saturation values $\mathcal{K}_{\text{sat}} \sim e^{S}$ where $S$ is the thermodynamic entropy.

Conversely, integrable systems exhibit Poissonian statistics with weaker level repulsion, which translates into significantly disordered Lanczos sequences with enhanced fluctuations. This irregularity reinforces state localization within the Krylov space, constraining operator spreading and consequently suppressing both the early-time growth rate and long-time saturation value of complexity \cite{Balasubramanian2022, Rabinovici2022, Espanol2023, Scialchi2024}. The connection between spectral statistics and Krylov space dynamics provides a mechanistic understanding of why chaotic systems explore Hilbert space more efficiently than their integrable counterparts.

\begin{figure}
\begin{center}
\includegraphics[width=0.95\linewidth]{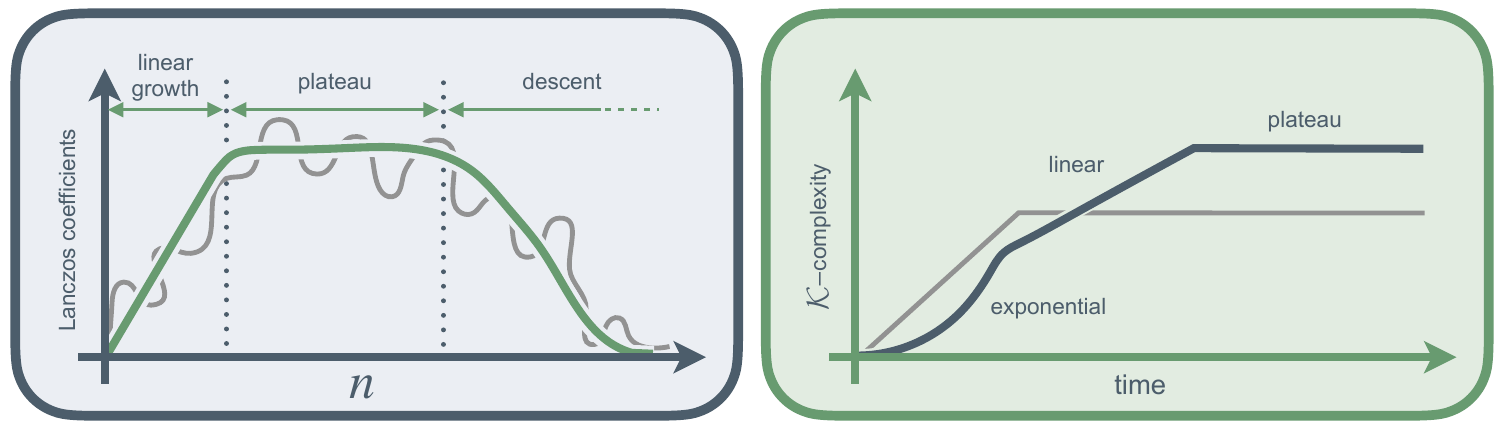}
\end{center}
\caption{\label{fig2:krylov}
Typical behavior of the off-diagonal Lanczos coefficients for both state and operator dynamics under chaotic Hamiltonians. We observe an initial linear growth regime, followed by a plateau, and finally a region where the coefficients gradually decay to zero. In the integrable case (gray line), the coefficients exhibit much stronger fluctuations. 
(b) Time evolution of the Krylov complexity for the chaotic case (black line) and the integrable case (gray line). At long times, the fluctuations of the Lanczos coefficients lead to a smaller asymptotic value of the Krylov complexity, while in the chaotic case an exponential growth regime is observed at short times.}
\end{figure}

While this framework has been numerically verified across diverse systems --including spin chains \cite{nation2021}, SYK models \cite{Rabinovici2022}, and quantum field theories \cite{caputa2022}-- important nuances remain. Recent studies indicate that the detailed behavior of Krylov complexity depends sensitively on both the initial state preparation and the specific operator chosen to generate the Krylov basis \cite{Espanol2023,Scialchi2024,pg2025dependence,craps2025multiseed}. These dependencies highlight that Krylov complexity provides complementary information to conventional measures, offering a more nuanced characterization of quantum chaos through the lens of Hilbert space dynamics. The consistent observation of these patterns across different physical platforms underscores Krylov complexity as a fundamental diagnostic tool for distinguishing chaotic from integrable dynamics in quantum systems.

A challenge arises when applying the standard Krylov method, typically defined for Hamiltonians, to systems described by a unitary evolution operator $U(t)$ [39, 40], common in areas like Quantum Reservoir Computing (QRC) [41-43] or kicked/Trotterized systems [44, 45].

For non-autonomous systems (e.g., kicked systems, Trotterized dynamics), the Lanczos algorithm is replaced by the \textit{Arnoldi algorithm}, which handles the unitary evolution operator $U$ directly. Given an initial state $|\psi_0\rangle$ and unitary $U$, the Krylov basis $\{|K_n\rangle\}$ is constructed via:
\[
b_n|K_n\rangle = U|K_{n-1}\rangle - \sum_{l=0}^{n-1} \langle K_l|U|K_{n-1}\rangle |K_l\rangle.
\]

The matrix elements of $U$ in the Krylov basis are,
\[
b_n = \langle K_n|U|K_{n-1}\rangle, \quad a_n = \langle K_n|U|K_n\rangle, \quad c_n = \langle K_0|U|K_n\rangle.
\]

The matrix $U$ exhibits a Hessenberg structure which, in the case of a fully chaotic system, has been conjectured to take on a particularly simple form, with unity elements along the subdiagonal and zeros elsewhere \cite{Suchsland2025}. Analytical evidence for this behavior has been demonstrated in a Floquet circuit. \cite{Suchsland2025}, and further numerical confirmation was provided in \cite{Scialchi2025}.

%%%%%Summary
\section{Summary and outlook}
\label{sec:summary}

The main goal of this tutorial has been to describe three complementary quantum diagnostics that capture, each in its own language, the intuition behind classical exponential sensitivity. Because unitary dynamics preserves Hilbert-space distances, no single quantity can play the role of a universal quantum Lyapunov exponent. Instead, the Loschmidt echo, OTOCs, and Krylov complexity provide operational notions of instability, irreversibility, and growth of complexity that can be compared across models and platforms.

The Loschmidt echo offers a direct notion of \emph{reversibility}: it quantifies how accurately an initial state can be reconstructed after an imperfect time-reversal protocol. Its decay regimes provide a concrete bridge to classical chaos in systems with semiclassical limits. In particular, the emergence of perturbation-independent decay rates tied to classical quantities (e.g., Lyapunov exponents or escape rates) constitutes one of the clearest manifestations of quantum--classical correspondence, while deviations from this behavior in mixed or many-body settings highlight the roles of coherence, conservation laws, and environmental couplings \cite{Gorin2006,Jacquod2009,DiegoScholar}.

OTOCs shift the emphasis from reversing trajectories to tracking \emph{operator growth}. They quantify how an initially simple operator becomes increasingly nonlocal under Heisenberg evolution and, therefore, how initially commuting operators fail to commute at later times. In semiclassical systems, the short-time behavior can mirror classical instability, but a key lesson is that early-time exponential growth of the squared commutator is not a universal indicator of chaos. More robust information is often contained in the intermediate- and late-time structure: the approach to the plateau, the decay rate of the OTOC in strongly chaotic systems, and the statistics of long-time fluctuations can all carry dynamical signatures, including clear distinctions between chaotic and integrable behavior \cite{otocScholarpedia,FortesPRE2019}. Recent experimental progress has also made OTOCs (and even higher-point generalizations) increasingly accessible, reinforcing their role as practical probes of scrambling in programmable quantum devices.

Krylov complexity provides a third viewpoint, rooted in the geometry of dynamics in a basis generated by the Hamiltonian (or Liouvillian) via Lanczos recursion. In this picture, evolution becomes a wavepacket propagating on an emergent one-dimensional chain with hopping amplitudes given by Lanczos coefficients. The resulting complexity measures how far the evolution explores this Krylov chain, offering an intrinsically dynamical diagnostic of ergodicity and structure in many-body systems \cite{Parker,nandy2025quantum,rabinovici2025krylov}. One appealing aspect of this approach is its constructive nature: rather than relying on a classical analogue, it organizes quantum evolution into a hierarchy of increasingly complex Krylov basis elements, making contact with spectral correlations, operator growth, and bounds on dynamical rates.

While the three diagnostics emphasize different operational viewpoints—echo recovery (Loschmidt), loss of commutativity and scrambling (OTOCs), and spreading in Krylov space (Krylov complexity)—they are not independent. Echo protocols can often be related to correlation functions and to variants of operator growth, and OTOCs can be interpreted as measuring the dynamical generation of nonlocal operator components that also drive spreading in Krylov space. Viewed together, these quantities provide a coherent picture: quantum dynamics can simultaneously remain unitary while exhibiting effective irreversibility, rapid delocalization of information, and growth of complexity, with the precise behavior controlled by chaos, integrability, disorder, conservation laws, and coupling to environments.

Several directions remain open and are likely to shape near-future developments. A first is improving the dictionary between intermediate-time relaxation (including possible quantum analogues of classical mixing rates) and the decay and saturation structure of OTOCs and Krylov observables in systems without a clean classical limit. A second is strengthening the experimental toolbox: scalable time-reversal protocols, randomized measurement strategies, and interference-based schemes are making increasingly sophisticated correlators accessible. Finally, it remains an important challenge to understand which aspects of “exponential sensitivity” are truly universal in quantum many-body dynamics and which depend on model-specific structures. The continued interplay between dynamical systems, quantum information, and many-body physics suggests that Loschmidt echoes, OTOCs, and Krylov complexity will remain central ingredients in this effort.
%%%%%%%%%%%

\bibliographystyle{JHEP}%
\bibliography{ref}

@article{Lorenz1963,
  author  = {Lorenz, Edward N.},
  title   = {Deterministic Nonperiodic Flow},
  journal = {Journal of the Atmospheric Sciences},
  volume  = {20},
  number  = {2},
  pages   = {130--141},
  year    = {1963},
  doi     = {10.1175/1520-0469(1963)020<0130:DNF>2.0.CO;2}
}

@article{bohigas,
	Author = {Bohigas, O. and Giannoni, M. J. and Schmit, C.},
	Doi = {10.1103/PhysRevLett.52.1},
	Issue = {1},
	Journal = {Phys. Rev. Lett.},
	Month = {Jan},
	Numpages = {0},
	Pages = {1--4},
	Publisher = {American Physical Society},
	Title = {Characterization of Chaotic Quantum Spectra and Universality of Level Fluctuation Laws},
	Url = {http://link.aps.org/doi/10.1103/PhysRevLett.52.1},
	Volume = {52},
	Year = {1984}}

@book{haake,
	Author = {F. Haake},
	Publisher = {{Springer-Verlag, Berlin}},
	Title = {Quantum Signatures of Chaos},
	Year = {1991}}

@article{guhr1998random,
  title={Random-matrix theories in quantum physics: common concepts},
  author={Guhr, Thomas and M{\"u}ller--Groeling, Axel and Weidenm{\"u}ller, Hans A},
  journal={Physics Reports},
  volume={299},
  number={4-6},
  pages={189--425},
  year={1998},
  publisher={Elsevier}
}

@article{Peres1984,
	Author = {Peres, Asher},
	Doi = {10.1103/PhysRevA.30.1610},
	Journal = {Phys. Rev. A},
	Month = {Oct},
	Number = {4},
	Numpages = {5},
	Pages = {1610},
	Publisher = {American Physical Society},
	Title = {Stability of quantum motion in chaotic and regular systems},
	Volume = {30},
	Year = {1984}}

@article{Jalabert2001,
	title = {Environment-Independent Decoherence Rate in Classically Chaotic Systems},
	author = {Jalabert, Rodolfo A. and Pastawski, Horacio M.},
	journal = {Phys. Rev. Lett.},
	volume = {86},
	issue = {12},
	pages = {2490--2493},
	numpages = {0},
	year = {2001},
	month = {Mar},
	publisher = {American Physical Society},
	doi = {10.1103/PhysRevLett.86.2490}
}

@article{Gorin2006,
  title={Scattering fidelity in elastodynamics},
  author={Gorin, Tomaz and Seligman, Thomas H and Weaver, Richard L},
  journal={Physical Review E},
  volume={73},
  number={1},
  pages={015202},
  year={2006},
  publisher={APS}
}

@article{Jacquod2009,
	Author = {\mbox{Ph.} Jacquod and C Petitjean},
	Journal = {Adv. Phys.},
	Number = {2},
	Pages = {67},
	Title = {Decoherence, entanglement and irreversibility in quantum dynamical systems with few degrees of freedom},
	Volume = {58},
	url={https://doi.org/10.1080/00018730902831009},
	Year = {2009}}

@article{DiegoScholar,
	Author = {A. Goussev and R. A. Jalabert and H. M. Pastawski and D. A. Wisniacki},
	Journal = {Scholarpedia},
	Pages = {11687},
	Title = {{L}oschmidt Echo},
	Volume = {7},
	url={http://www.scholarpedia.org/article/Loschmidt_echo},
	Year = {2012}}

@article{Prosen2002,
	Author = {Tomaz Prosen and Marko {\v Z}nidari{\v c}},
	Journal = {J. Phys. A: Math. Gen.},
	Pages = {1455},
	Title = {Stability of quantum motion and correlation decay},
	Volume = {35},
	Year = {2002}}

@article{Cucc2003,
	Author = {FM Cucchietti and DAR Dalvit and JP Paz and WH Zurek},
	Journal = {Phys. Rev. Lett.},
	Number = {21},
	Pages = {210403--210403},
	Title = {Decoherence and the {L}oschmidt echo},
	Volume = {91},
	Year = {2003}}

@article{Quan2006,
	Author = {HT Quan and Z Song and XF Liu and P Zanardi and CP Sun},
	Journal = prl,
	Pages = {140604},
	Title = {Decay of {L}oschmidt Echo Enhanced by Quantum Criticality},
	Volume = {96},
	Year = {2006}}

@article{larkin1969quasiclassical,
  title={Quasiclassical method in the theory of superconductivity},
  author={Larkin, AI and Ovchinnikov, Yu N},
  journal={Sov. Phys. JETP},
  volume={28},
  number={6},
  pages={1200--1205},
  url={http://www.jetp.ras.ru/cgi-bin/dn/e_028_06_1200.pdf},
  year={1969}
}

@article{shenker2014black,
  title={Black holes and the butterfly effect},
  author={Shenker, Stephen H and Stanford, Douglas},
  journal={J. High Energy Phys.},
  volume={2014},
  number={3},
  pages={67},
  year={2014},
  url={https://link.springer.com/article/10.1007/JHEP03(2014)067},
  publisher={Springer}
}

@article{maldacena2016,
    title={Out-of-time-order correlators in quantum chaos},
    author={Maldacena, Juan and Shenker, Stephen H. and Stanford, Douglas},
    journal={Journal of High Energy Physics},
    volume={2016},
    number={8},
    pages={1--24},
    year={2016},
    publisher={Springer}
}

@article{swingle2018,
  title = {Unscrambling the physics of out-of-time-order correlators},
  author = {Swingle, Brian},
  journal = {Nature Physics},
  volume = {14},
  pages = {988},
  year = {2018},
  doi = {10.1038/s41567-018-0295-5},
}

@article{otocScholarpedia,
author={Garc\'ia-Mata, I. and  Jalabert, R.A.  and Wisniacki, D.A.},
Title={Out-of-time-order correlations and quantum chaos},
year = {2023},
journal= {Scholarpedia},
volume= {18},
number= {4},
pages= {55237},
DOI = {10.4249/scholarpedia.55237},
NOTA = {revisión \#204529}
}

@book{kitaev2015simple,
  title={A simple model of quantum holography, talk given at KITP Program: Entanglement in Strongly-Correlated Quantum Matter},
  author={Kitaev, Alexei},
  Publisher={USA April},
  volume={7},
  year={2015}
}

@article{Parker,
	author = {Parker, Daniel E. and Cao, Xiangyu and Avdoshkin, Alexander and Scaffidi, Thomas and Altman, Ehud},
	doi = {10.1103/PhysRevX.9.041017},
	issue = {4},
	journal = {Phys. Rev. X},
	month = {Oct},
	numpages = {29},
	pages = {041017},
	publisher = {American Physical Society},
	title = {A Universal Operator Growth Hypothesis},
	url = {https://link.aps.org/doi/10.1103/PhysRevX.9.041017},
	volume = {9},
	year = {2019}}

@article{nandy2025quantum,
  title={Quantum dynamics in Krylov space: Methods and applications},
  author={Nandy, Pratik and Matsoukas-Roubeas, Apollonas S and Mart{\'\i}nez-Azcona, Pablo and Dymarsky, Anatoly and del Campo, Adolfo},
  journal={Physics Reports},
  volume={1125},
  pages={1--82},
  year={2025},
  publisher={Elsevier}
}

@article{rabinovici2025krylov,
  title={Krylov Complexity},
  author={Rabinovici, Eliezer and S{\'a}nchez-Garrido, Adri{\'a}n and Shir, Ruth and Sonner, Julian},
  journal={arXiv preprint arXiv:2507.06286},
  year={2025}
}

@article{goirin_PhysRep2006,
title = {Dynamics of {L}oschmidt echoes and fidelity decay},
journal = {Physics Reports},
volume = {435},
number = {2},
pages = {33-156},
year = {2006},
issn = {0370-1573},
doi = {https://doi.org/10.1016/j.physrep.2006.09.003},
url = {https://www.sciencedirect.com/science/article/pii/S0370157306003310},
author = {Thomas Gorin and Tomaž Prosen and Thomas H. Seligman and Marko Žnidarič},
}

@article{wang2021microscope,
  title = {Microscope for quantum dynamics with {P}lanck cell resolution},
  author = {Wang, Zhenduo and Feng, Jiajin and Wu, Biao},
  journal = {Phys. Rev. Research},
  volume = {3},
  issue = {3},
  pages = {033239},
  numpages = {9},
  year = {2021},
  month = {Sep},
  publisher = {American Physical Society},
  doi = {10.1103/PhysRevResearch.3.033239},
  url = {https://link.aps.org/doi/10.1103/PhysRevResearch.3.033239}
}

@article{dematos1995quantization,
  title={Quantization of Anosov maps},
  author={Dematos, M Basilio and Dealmeida, AM Ozorio},
  journal={Annals of Physics},
  volume={237},
  number={1},
  pages={46--65},
  year={1995},
  publisher={Elsevier}
}

@book{Gutzwiller1990,
	title = {Chaos in Classical and Quantum Mechanics},
	author = {Gutzwiller, Martin C.},
	series = {Interdisciplinary Applied Mathematics},
	volume = {1},
	publisher = {Springer-Verlag},
	address = {New York},
	year = {1990},
	doi = {10.1007/978-1-4612-0983-6}
}

@article{Jacquod2001,
	title = {Golden rule decay versus Lyapunov decay of the quantum Loschmidt echo},
	author = {Jacquod, Ph. and Silvestrov, P. G. and Beenakker, C. W. J.},
	journal = {Phys. Rev. E},
	volume = {64},
	issue = {5},
	pages = {055203},
	numpages = {4},
	year = {2001},
	month = {Oct},
	publisher = {American Physical Society},
	doi = {10.1103/PhysRevE.64.055203}
}

@article{Vanicek2004,
	title = {Dephasing representation: Employing the shadowing theorem to calculate quantum correlation functions},
	author = {Vanicek, Jiri},
	journal = {Phys. Rev. E},
	volume = {70},
	issue = {5},
	pages = {055201},
	numpages = {4},
	year = {2004},
	month = {Nov},
	publisher = {American Physical Society},
	doi = {10.1103/PhysRevE.70.055201}
}

@article{Vanicek2006,
	title = {Dephasing representation of quantum fidelity for general pure and mixed states},
	author = {Vanicek, Jiri},
	journal = {Phys. Rev. E},
	volume = {73},
	issue = {4},
	pages = {046204},
	numpages = {10},
	year = {2006},
	month = {Apr},
	publisher = {American Physical Society},
	doi = {10.1103/PhysRevE.73.046204}
}

@article{Gorin2004,
	title = {A random matrix formulation of fidelity decay},
	author = {Gorin, T. and Prosen, T. and Seligman, T. H.},
	journal = {New J. Phys.},
	volume = {6},
	number = {1},
	pages = {20},
	year = {2004},
	month = {Feb},
	publisher = {IOP Publishing},
	doi = {10.1088/1367-2630/6/1/020}
}

@article{Stoeckmann2004,
	title = {Recovery of the fidelity amplitude for the Gaussian ensembles},
	author = {Stoeckmann, H.-J. and Sch\"afer, R.},
	journal = {New J. Phys.},
	volume = {6},
	pages = {199},
	year = {2004},
	month = {Dec},
	publisher = {IOP Publishing},
	doi = {10.1088/1367-2630/6/1/199}
}

@article{Kohler2008,
	title = {Surprising Relations between Parametric Level Correlations and Fidelity Decay},
	author = {Kohler, H. and Smolyarenko, I. E. and Pineda, C. and Guhr, T. and Leyvraz, F. and Seligman, T. H.},
	journal = {Phys. Rev. Lett.},
	volume = {100},
	issue = {19},
	pages = {190404},
	numpages = {4},
	year = {2008},
	month = {May},
	publisher = {American Physical Society},
	doi = {10.1103/PhysRevLett.100.190404}
}

@article{Cucchietti2002,
	title = {Measuring the Lyapunov exponent using quantum mechanics},
	author = {Cucchietti, F. M. and Lewenkopf, C. H. and Mucciolo, E. R. and Pastawski, H. M. and Vallejos, R. O.},
	journal = {Phys. Rev. E},
	volume = {65},
	issue = {4},
	pages = {046209},
	numpages = {7},
	year = {2002},
	month = {Mar},
	publisher = {American Physical Society},
	doi = {10.1103/PhysRevE.65.046209}
}

@article{DeRaedt1996,
	title = {Computer simulation of quantum phenomena in nanoscale devices},
	author = {De Raedt, H.},
	journal = {Annu. Rev. Comput. Phys.},
	volume = {4},
	pages = {107--146},
	year = {1996},
	publisher = {World Scientific},
	doi = {10.1142/9789812816152_0004}
}

@article{TalEzer1984,
	title = {An accurate and efficient scheme for propagating the time dependent Schr\"{o}dinger equation},
	author = {Tal-Ezer, H. and Kosloff, R.},
	journal = {J. Chem. Phys.},
	volume = {81},
	number = {9},
	pages = {3967--3971},
	year = {1984},
	month = {Nov},
	publisher = {AIP Publishing},
	doi = {10.1063/1.448136}
}

@article{gutierrez2009,
	Author = {Guti\'errez, Martha and Goussev, Arseni},
	Doi = {10.1103/PhysRevE.79.046211},
	Journal = {Phys. Rev. E},
	Month = {Apr},
	Number = {4},
	Numpages = {5},
	Pages = {046211},
	Publisher = {American Physical Society},
	Title = {Long-time saturation of the {L}oschmidt echo in quantum chaotic billiards},
	Volume = {79},
	Year = {2009}}

@article{Goussev2008,
	Author = {Arseni Goussev and Daniel Waltner and Klaus Richter and Rodolfo A Jalabert},
	Journal = {New J. Phys.},
	Pages = {093010},
	Title = {Loschmidt echo for local perturbations: non-monotonic cross-over from the Fermi-golden-rule to the escape-rate regime},
	Volume = {10},
	Year = {2008}}

@article{goussev2008loschmidt,
  title={Loschmidt echo for local perturbations: non-monotonic cross-over from the Fermi-golden-rule to the escape-rate regime},
  author={Goussev, Arseni and Waltner, Daniel and Richter, Klaus and Jalabert, Rodolfo A},
  journal={New Journal of Physics},
  volume={10},
  number={9},
  pages={093010},
  year={2008},
  publisher={IOP Publishing}
}

@article{ares2009loschmidt,
  title={Loschmidt echo and the local density of states},
  author={Ares, Natalia and Wisniacki, Diego A},
  journal={Physical Review E—Statistical, Nonlinear, and Soft Matter Physics},
  volume={80},
  number={4},
  pages={046216},
  year={2009},
  publisher={APS}
}

@article{Weinstein2005,
	Author = {Yaakov Weinstein and C Hellberg},
	Journal = {Phys. Rev. E},
	Pages = {016209},
	Title = {Quantum fidelity decay in quasi-integrable systems},
	Volume = {71},
	Year = {2005}}

@article{jacquod2003anomalous,
  title={Anomalous power law of quantum reversibility forclassically regular dynamics},
  author={Jacquod, Ph and Adagideli, I and Beenakker, Carlo WJ},
  journal={Europhysics letters},
  volume={61},
  number={6},
  pages={729},
  year={2003},
  publisher={IOP Publishing}
}

@article{prosen2003quantum,
  title={Quantum freeze of fidelity decay for a class of integrable dynamics},
  author={Prosen, Toma{\v{z}} and {\v{Z}}nidari{\v{c}}, Marko},
  journal={New Journal of Physics},
  volume={5},
  number={1},
  pages={109},
  year={2003},
  publisher={IOP Publishing}
}

@article{Hahn1950,
  title={Spin echoes},
  author={Hahn, Erwin L},
  journal={Physical Review},
  volume={80},
  number={4},
  pages={580--594},
  year={1950},
  publisher={APS}
}

@article{Rhim1971,
  title={Time-reversal experiments in dipolar-coupled spin systems},
  author={Rhim, Won-Kyu and Pines, Alexander and Waugh, John S},
  journal={Physical Review B},
  volume={3},
  number={3},
  pages={684--696},
  year={1971},
  publisher={APS}
}

@article{Levstein1998,
  title={Attenuation of polarization echoes in nuclear magnetic resonance: A study of the emergence of dynamical irreversibility in many-body quantum systems},
  author={Levstein, Patricia R and Usaj, Gonzalo and Pastawski, Horacio M},
  journal={The Journal of Chemical Physics},
  volume={108},
  number={7},
  pages={2718--2724},
  year={1998},
  publisher={AIP Publishing}
}

@article{Usaj1998,
  title={Gaussian to exponential crossover in the attenuation of polarization echoes in NMR},
  author={Usaj, Gonzalo and Pastawski, Horacio M and Levstein, Patricia R},
  journal={Molecular Physics},
  volume={95},
  number={6},
  pages={1229--1236},
  year={1998},
  publisher={Taylor \& Francis}
}

@article{Pastawski2000,
  title={A nuclear magnetic resonance answer to the Boltzmann-Loschmidt controversy?},
  author={Pastawski, Horacio M and Levstein, Patricia R and Usaj, Gonzalo and Raya, Jesus and Hirschinger, J{\"o}rg},
  journal={Physica A: Statistical Mechanics and its Applications},
  volume={283},
  number={3-4},
  pages={166--170},
  year={2000},
  publisher={Elsevier}
}

@article{Schafer2005,
  title={Experimental verification of fidelity decay: From perturbative to Fermi golden rule regime},
  author={Sch{\"a}fer, Rudolf and St{\"o}ckmann, Hans-J{\"u}rgen and Gorin, Tomaz and Seligman, Thomas H},
  journal={Physical Review Letters},
  volume={95},
  number={18},
  pages={184102},
  year={2005},
  publisher={APS}
}

@article{Stockmann2005,
  title={Fidelity recovery in chaotic systems and the Debye-Waller factor},
  author={St{\"o}ckmann, Hans-J{\"u}rgen and Sch{\"a}fer, Rudolf},
  journal={Physical Review Letters},
  volume={94},
  number={24},
  pages={244101},
  year={2005},
  publisher={APS}
}

@article{Kober2011,
  title={Fidelity decay for local perturbations: Microwave evidence for oscillating decay exponents},
  author={Kober, B and Kuhl, U and St{\"o}ckmann, H-J and Goussev, A and Richter, K},
  journal={Physical Review E},
  volume={83},
  number={1},
  pages={016214},
  year={2011},
  publisher={APS}
}

@article{Lobkis2003,
  title={Coda-wave interferometry in finite solids: Recovery of P-to-S conversion rates in an elastodynamic billiard},
  author={Lobkis, Oleg I and Weaver, Richard L},
  journal={Physical Review Letters},
  volume={90},
  number={25},
  pages={254302},
  year={2003},
  publisher={APS}
}

@article{Andersen2006,
  title={Decay of quantum correlations in atom optics billiards with chaotic and mixed dynamics},
  author={Andersen, MF and Kaplan, A and Gr{\"u}nzweig, T and Davidson, N},
  journal={Physical Review Letters},
  volume={97},
  number={10},
  pages={104102},
  year={2006},
  publisher={APS}
}

@article{Wu2009,
  title={Observation of saturation of fidelity decay with an atom interferometer},
  author={Wu, Saijun and Tonyushkin, Alexey and Prentiss, Mara G},
  journal={Physical Review Letters},
  volume={103},
  number={3},
  pages={034101},
  year={2009},
  publisher={APS}
}

@article{Ullah2011,
  title={Experimental observation of Loschmidt time reversal of a quantum chaotic system},
  author={Ullah, A and Hoogerland, MD},
  journal={Physical Review E},
  volume={83},
  number={4},
  pages={046218},
  year={2011},
  publisher={APS}
}

@article{Mart2008,
  title={Cooling by time reversal of atomic matter waves},
  author={Martin, J{\'e}r{\'e}my and Georgeot, Bertrand and Shepelyansky, Dima L},
  journal={Physical Review Letters},
  volume={100},
  number={4},
  pages={044106},
  year={2008},
  publisher={APS}
}

@article{Fink1999,
  title={Time-reversed acoustics},
  author={Fink, Mathias},
  journal={Scientific American},
  volume={281},
  number={5},
  pages={91--97},
  year={1999},
  publisher={Scientific American, a division of Nature America, Inc.}
}

@article{Calvo2008,
  title={Semiclassical theory of time-reversal focusing},
  author={Calvo, Hern{\'a}n L and Jalabert, Rodolfo A and Pastawski, Horacio M},
  journal={Physical Review Letters},
  volume={101},
  number={24},
  pages={240403},
  year={2008},
  publisher={APS}
}

@article{Deutsch2018,
  title={Eigenstate thermalization hypothesis},
  author={Deutsch, Joshua M},
  journal={Reports on Progress in Physics},
  volume={81},
  number={8},
  pages={082001},
  year={2018},
  publisher={IOP Publishing}
}

@article{DAlessio2016,
  title={From quantum chaos and eigenstate thermalization to statistical mechanics and thermodynamics},
  author={D'Alessio, Luca and Kafri, Yariv and Polkovnikov, Anatoli and Rigol, Marcos},
  journal={Advances in Physics},
  volume={65},
  number={3},
  pages={239--362},
  year={2016},
  publisher={Taylor \& Francis}
}

@article{Heyl2018,
  title={Dynamical quantum phase transitions: a review},
  author={Heyl, Markus},
  journal={Reports on Progress in Physics},
  volume={81},
  number={5},
  pages={054001},
  year={2018},
  publisher={IOP Publishing}
}

@article{Calabrese2016,
  title={Quantum quenches in 1+1 dimensional conformal field theories},
  author={Calabrese, Pasquale and Cardy, John},
  journal={Journal of Statistical Mechanics: Theory and Experiment},
  volume={2016},
  number={6},
  pages={064001},
  year={2016},
  publisher={IOP Publishing}
}

@article{Altman2018,
  title={Many-body localization},
  author={Altman, Ehud},
  journal={Annalen der Physik},
  volume={530},
  number={7},
  pages={1700346},
  year={2018},
  publisher={Wiley Online Library}
}

@article{Nandkishore2015,
  title={Many-body localization and thermalization in quantum statistical mechanics},
  author={Nandkishore, Rahul and Huse, David A},
  journal={Annual Review of Condensed Matter Physics},
  volume={6},
  pages={15--38},
  year={2015},
  publisher={Annual Reviews}
}

@article{Brown2016,
  title={Loschmidt echo and the many-body light cone},
  author={Brown, Adam R and Roberts, Daniel A and Swingle, Brian and Zhao, Yue and Zhang, Ying},
  journal={Physical Review Letters},
  volume={116},
  number={22},
  pages={220601},
  year={2016},
  publisher={APS}
}

@article{Cotler2017,
  title={Black holes and random matrices},
  author={Cotler, Jordan S and Gur-Ari, Guy and Hanada, Masanori and Polchinski, Joseph and Saad, Phil and Shenker, Stephen H and Stanford, Douglas and Streicher, Alexandre and Tezuka, Masaki},
  journal={Journal of High Energy Physics},
  volume={2017},
  number={5},
  pages={1--54},
  year={2017},
  publisher={Springer}
}

@article{Mi2022,
  title={Information scrambling in quantum circuits},
  author={Mi, Xiao and Ippoliti, Matteo and Quintana, Chris and Greene, Ami and Chen, Zijun and Ho, Alan and Choi, Seung and Collins, Ravi and Awschalom, David and Schuster, David and others},
  journal={Science},
  volume={376},
  number={6593},
  pages={abg5029},
  year={2022},
  publisher={American Association for the Advancement of Science}
}

@article{Zhang2017,
  title={Observation of a many-body dynamical phase transition with a 53-qubit quantum simulator},
  author={Zhang, J and Pagano, G and Hess, PW and Kyprianidis, A and Becker, P and Kaplan, H and Gorshkov, AV and Gong, Z-X and Monroe, C},
  journal={Nature},
  volume={551},
  number={7682},
  pages={601--604},
  year={2017},
  publisher={Nature Publishing Group}
}

@article{yan_zurek2020,
  title = {Information Scrambling and {L}oschmidt Echo},
  author = {Yan, Bin and Cincio, Lukasz and Zurek, Wojciech H.},
  journal = {Phys. Rev. Lett.},
  volume = {124},
  issue = {16},
  pages = {160603},
  numpages = {6},
  year = {2020},
  month = {Apr},
  publisher = {American Physical Society},
  doi = {10.1103/PhysRevLett.124.160603},
  url = {https://link.aps.org/doi/10.1103/PhysRevLett.124.160603}
}

@article{liao2018nonlinear,
  title = {Nonlinear sigma model approach to many-body quantum chaos: Regularized and unregularized out-of-time-ordered correlators},
  author = {Liao, Yunxiang and Galitski, Victor},
  journal = {Phys. Rev. B},
  volume = {98},
  issue = {20},
  pages = {205124},
  numpages = {21},
  year = {2018},
  month = {Nov},
  publisher = {American Physical Society},
  doi = {10.1103/PhysRevB.98.205124},
  url = {https://link.aps.org/doi/10.1103/PhysRevB.98.205124}
}

@article{romero2019,
  title = {Regularization dependence of the OTOC. {W}hich {L}yapunov spectrum is the physical one?},
  author = {Romero-Berm\'udez, Aurelio and Schalm, Koenraad and Scopelliti, Vincenzo},
  journal = {Journal of High Energy Physics},
  volume = {2019},
  pages = {107},
  year = {2019},
  month = {July},
  publisher = {Springer},
  doi = {10.1007/JHEP07(2019)107},
  url = {https://doi.org/10.1007/JHEP07(2019)107}
}

@article{sahu2020,
  title = {Information scrambling at finite temperature in local quantum systems},
  author = {Sahu, Subhayan and Swingle, Brian},
  journal = {Phys. Rev. B},
  volume = {102},
  issue = {18},
  pages = {184303},
  numpages = {22},
  year = {2020},
  month = {Nov},
  publisher = {American Physical Society},
  doi = {10.1103/PhysRevB.102.184303},
  url = {https://link.aps.org/doi/10.1103/PhysRevB.102.184303}
}

@article{shenker2015stringy,
  title={Stringy effects in scrambling},
  author={Shenker, Stephen H and Stanford, Douglas},
  journal={J. High Energy Phys.},
  volume={2015},
  number={5},
  pages={132},
  year={2015},
  publisher={Springer}
}

@article{LiebRobinson72,
  title={Lieb-{R}obinson Bounds and the Exponential Clustering Theorem},
  author={Lieb, E. H. and Robinson, D. W.},
  journal={Commun. Math. Phys.},
  volume={28},
  pages={251},
  url={https://link.springer.com/article/10.1007/s00220-006-1556-1},
  year={1972}
}

@article{roberts_swingle_2016,
  title = {{L}ieb-{R}obinson Bound and the Butterfly Effect in Quantum Field Theories},
  author = {Roberts, Daniel A. and Swingle, Brian},
  journal = {Phys. Rev. Lett.},
  volume = {117},
  issue = {9},
  pages = {091602},
  numpages = {6},
  year = {2016},
  month = {Aug},
  publisher = {American Physical Society},
  doi = {10.1103/PhysRevLett.117.091602},
  url = {https://link.aps.org/doi/10.1103/PhysRevLett.117.091602}
}

@article{Xu2019Butterfly,
  title = {Butterfly effect in interacting Aubry-Andre model: Thermalization, slow scrambling, and many-body localization},
  author = {Xu, Shenglong and Li, Xiao and Hsu, Yi-Ting and Swingle, Brian and Das Sarma, S.},
  journal = {Phys. Rev. Res.},
  volume = {1},
  issue = {3},
  pages = {032039},
  numpages = {6},
  year = {2019},
  month = {Dec},
  publisher = {American Physical Society},
  doi = {10.1103/PhysRevResearch.1.032039},
  url = {https://link.aps.org/doi/10.1103/PhysRevResearch.1.032039}
}

@article{Xu2019Locality,
  title = {Locality, Quantum Fluctuations, and Scrambling},
  author = {Xu, Shenglong and Swingle, Brian},
  journal = {Phys. Rev. X},
  volume = {9},
  issue = {3},
  pages = {031048},
  numpages = {21},
  year = {2019},
  month = {Sep},
  publisher = {American Physical Society},
  doi = {10.1103/PhysRevX.9.031048},
  url = {https://link.aps.org/doi/10.1103/PhysRevX.9.031048}
}

@article{swingle2020,
  title = {Accessing scrambling using matrix product operators
},
  author = {Xu, Shenglong and Swingle, Brian},
  journal = {Nature Physics},
  volume = {16},
  pages = {199},
  year = {2020},
  doi = {10.1038/s41567-019-0712-4},
}

@article{kurchan2018quantum,
  title={Quantum bound to chaos and the semiclassical limit},
  author={Kurchan, Jorge},
  journal={Journal of Statistical Physics},
  volume={171},
  number={6},
  pages={965--979},
  year={2018},
  publisher={Springer}
}

@article{schmitt2018,
  title = {Irreversible dynamics in quantum many-body systems},
  author = {Schmitt, Markus and Kehrein, Stefan},
  journal = {Phys. Rev. B},
  volume = {98},
  issue = {18},
  pages = {180301},
  numpages = {6},
  year = {2018},
  month = {Nov},
  publisher = {American Physical Society},
  doi = {10.1103/PhysRevB.98.180301},
  url = {https://link.aps.org/doi/10.1103/PhysRevB.98.180301}
}

@article{schmitt2019,
  title = {Semiclassical echo dynamics in the Sachdev-Ye-Kitaev model},
  author = {Schmitt, Markus and Sels, Dries and Kehrein, Stefan and Polkovnikov, Anatoli},
  journal = {Phys. Rev. B},
  volume = {99},
  issue = {13},
  pages = {134301},
  numpages = {10},
  year = {2019},
  month = {Apr},
  publisher = {American Physical Society},
  doi = {10.1103/PhysRevB.99.134301},
  url = {https://link.aps.org/doi/10.1103/PhysRevB.99.134301}
}

@article{Sanchez2021,
  title = {Emergent decoherence induced by quantum chaos in a many-body system: A {L}oschmidt echo observation through {NMR}},
  author = {S\'anchez, C. M. and Chattah, A. K. and Pastawski, H. M.},
  journal = {Phys. Rev. A},
  volume = {105},
  issue = {5},
  pages = {052232},
  numpages = {11},
  year = {2022},
  month = {May},
  publisher = {American Physical Society},
  doi = {10.1103/PhysRevA.105.052232},
  url = {https://link.aps.org/doi/10.1103/PhysRevA.105.052232}
}

@article{rammensee2018,
  title = {Many-Body Quantum Interference and the Saturation of Out-of-Time-Order Correlators},
  author = {Rammensee, Josef and Urbina, Juan Diego and Richter, Klaus},
  journal = {Phys. Rev. Lett.},
  volume = {121},
  issue = {12},
  pages = {124101},
  numpages = {6},
  year = {2018},
  month = {Sep},
  publisher = {American Physical Society},
  doi = {10.1103/PhysRevLett.121.124101},
  url = {https://link.aps.org/doi/10.1103/PhysRevLett.121.124101}
}

@article{FortesPRE2019,
  title = {Gauging classical and quantum integrability through out-of-time-ordered correlators},
  author = {Fortes, Emiliano M. and Garc\'{\i}a-Mata, Ignacio and Jalabert, Rodolfo A. and Wisniacki, Diego A.},
  journal = {Phys. Rev. E},
  volume = {100},
  issue = {4},
  pages = {042201},
  numpages = {13},
  year = {2019},
  month = {Oct},
  publisher = {American Physical Society},
  doi = {10.1103/PhysRevE.100.042201},
  url = {https://link.aps.org/doi/10.1103/PhysRevE.100.042201}
}

@article{DimaScholar_Ehrenfest,
	Author = {D. L. Shepelyansky},
	Journal = {Scholarpedia},
	Pages = {55031},
	Title = {Ehrenfest time and chaos},
	Volume = {15(9)},
	Year = {2020}}

@article{polchinski2015,
  title={Chaos in the black hole {S}-matrix},
  author={Polchinski, Joseph},
  journal={arXiv:1505.08108},
  doi={ https://doi.org/10.48550/arXiv.1505.08108},
  url={ https://arxiv.org/abs/1505.08108},
  year={2015}
}

@article{jahnke2019chaos,
  title={On the chaos bound in rotating black holes},
  author={Jahnke, Viktor and Kim, Keun-Young and Yoon, Junggi},
  journal={Journal of High Energy Physics},
  volume={2019},
  number={5},
  pages={1--35},
  year={2019},
  url={https://link.springer.com/article/10.1007/JHEP05(2019)037},
  publisher={Springer}
}

@article{sekino2008fast,
  title={Fast scramblers},
  author={Sekino, Yasuhiro and Susskind, Leonard},
  journal={Journal of High Energy Physics},
  volume={2008},
  number={10},
  pages={065},
  year={2008},
  url={https://iopscience.iop.org/article/10.1088/1126-6708/2008/10/065/meta?casa_token=4viSmsdGzy4AAAAA:ZQvFPUAAbqUl0X2Bj7VKI4dQnNIOYbgtSg5oxZxOhmuve_WYiriYBhzLNdTDQiE-YnsjEvRInxHen7U},
  publisher={IOP Publishing}
}

@article{hosur2016chaos,
  title={Chaos in quantum channels},
  author={Hosur, Pavan and Qi, Xiao-Liang and Roberts, Daniel A and Yoshida, Beni},
  journal={Journal of High Energy Physics},
  volume={2016},
  number={2},
  pages={1--49},
  year={2016},
  url={https://link.springer.com/article/10.1007/JHEP02(2016)004#citeas},
  publisher={Springer}
}

@article{JGMW2018,
  title = {Semiclassical theory of out-of-time-order correlators for low-dimensional classically chaotic systems},
  author = {Jalabert, Rodolfo A. and Garc\'{\i}a-Mata, Ignacio and Wisniacki, Diego A.},
  journal = {Phys. Rev. E},
  volume = {98},
  issue = {6},
  pages = {062218},
  numpages = {12},
  year = {2018},
  month = {Dec},
  publisher = {American Physical Society},
  doi = {10.1103/PhysRevE.98.062218},
  url = {https://link.aps.org/doi/10.1103/PhysRevE.98.062218}
}

@article{Rozenbaum2017,
  title = {Lyapunov Exponent and Out-of-Time-Ordered Correlator's Growth Rate in a Chaotic System},
  author = {Rozenbaum, Efim B. and Ganeshan, Sriram and Galitski, Victor},
  journal = {Phys. Rev. Lett.},
  volume = {118},
  issue = {8},
  pages = {086801},
  numpages = {5},
  year = {2017},
  month = {Feb},
  publisher = {American Physical Society},
  doi = {10.1103/PhysRevLett.118.086801},
  url = {https://link.aps.org/doi/10.1103/PhysRevLett.118.086801}
}

@article{OTOC_gato_PRL,
  title = {Chaos Signatures in the Short and Long Time Behavior of the Out-of-Time Ordered Correlator},
  author = {Garc\'{\i}a-Mata, Ignacio and Saraceno, Marcos and Jalabert, Rodolfo A. and Roncaglia, Augusto J. and Wisniacki, Diego A.},
  journal = {Phys. Rev. Lett.},
  volume = {121},
  issue = {21},
  pages = {210601},
  numpages = {5},
  year = {2018},
  month = {Nov},
  publisher = {American Physical Society},
  doi = {10.1103/PhysRevLett.121.210601},
  url = {https://link.aps.org/doi/10.1103/PhysRevLett.121.210601}
}

@article{lakshminarayan2019out,
  title = {Out-of-time-ordered correlator in the quantum bakers map and truncated unitary matrices},
  author = {Lakshminarayan, Arul},
  journal = {Phys. Rev. E},
  volume = {99},
  issue = {1},
  pages = {012201},
  numpages = {9},
  year = {2019},
  month = {Jan},
  publisher = {American Physical Society},
  doi = {10.1103/PhysRevE.99.012201},
  url = {https://link.aps.org/doi/10.1103/PhysRevE.99.012201}
}

@article{morita2021extracting,
         title = {Extracting Classical {L}yapunov Exponent from One-Dimensional Quantum Mechanics},
         author = {Takeshi Morita },
         year = {2021},
         journal = {2105.09603v2},
         archivePrefix = {arXiv},
          url={https://arxiv.org/abs/2105.09603},
         primaryClass ={hep-th}
        }

@article{chavez2019quantum,
  title = {Quantum and Classical {L}yapunov Exponents in Atom-Field Interaction Systems},
  author = {Ch\'avez-Carlos, Jorge and L\'opez-del-Carpio, B. and Bastarrachea-Magnani, Miguel A. and Str\'ansk\'y, Pavel and Lerma-Hern\'andez, Sergio and Santos, Lea F. and Hirsch, Jorge G.},
  journal = {Phys. Rev. Lett.},
  volume = {122},
  issue = {2},
  pages = {024101},
  numpages = {7},
  year = {2019},
  month = {Jan},
  publisher = {American Physical Society},
  doi = {10.1103/PhysRevLett.122.024101},
  url = {https://link.aps.org/doi/10.1103/PhysRevLett.122.024101}
}

@article{ali2020chaos,
  title = {Chaos and complexity in quantum mechanics},
  author = {Ali, Tibra and Bhattacharyya, Arpan and Haque, S. Shajidul and Kim, Eugene H. and Moynihan, Nathan and Murugan, Jeff},
  journal = {Phys. Rev. D},
  volume = {101},
  issue = {2},
  pages = {026021},
  numpages = {11},
  year = {2020},
  month = {Jan},
  publisher = {American Physical Society},
  doi = {10.1103/PhysRevD.101.026021},
  url = {https://link.aps.org/doi/10.1103/PhysRevD.101.026021}
}

@misc{Lorenz1972,
  author       = {Lorenz, Edward N.},
  title        = {Predictability: Does the Flap of a Butterfly's Wings in Brazil Set Off a Tornado in Texas?},
  howpublished = {Talk presented at the 139th Meeting of the American Association for the Advancement of Science (AAAS)},
  address      = {Washington, D.C.},
  year         = {1972}
}

@article{Roberts2015,
  author  = {Roberts, Daniel A. and Stanford, Douglas and Susskind, Leonard},
  title   = {Localized shocks},
  journal = {Journal of High Energy Physics},
  volume  = {2015},
  number  = {3},
  pages   = {51},
  year    = {2015},
  doi     = {10.1007/JHEP03(2015)051}
}

@article{Swingle2017,
  title = {Slow scrambling in disordered quantum systems},
  author = {Swingle, Brian and Chowdhury, Debanjan},
  journal = {Phys. Rev. B},
  volume = {95},
  issue = {6},
  pages = {060201},
  numpages = {6},
  year = {2017},
  month = {Feb},
  publisher = {American Physical Society},
  doi = {10.1103/PhysRevB.95.060201},
  url = {https://link.aps.org/doi/10.1103/PhysRevB.95.060201}
}

@article{craps2020lyapunov,
  title = {Lyapunov growth in quantum spin chains},
  author = {Craps, Ben and De Clerck, Marine and Janssens, Djunes and Luyten, Vincent and Rabideau, Charles},
  journal = {Phys. Rev. B},
  volume = {101},
  issue = {17},
  pages = {174313},
  numpages = {17},
  year = {2020},
  month = {May},
  publisher = {American Physical Society},
  doi = {10.1103/PhysRevB.101.174313},
  url = {https://link.aps.org/doi/10.1103/PhysRevB.101.174313}
}

@article{Motrunich2018,
  title = {Out-of-time-ordered correlators in a quantum {I}sing chain},
  author = {Lin, Cheng-Ju and Motrunich, Olexei I.},
  journal = {Phys. Rev. B},
  volume = {97},
  issue = {14},
  pages = {144304},
  numpages = {17},
  year = {2018},
  month = {Apr},
  publisher = {American Physical Society},
  doi = {10.1103/PhysRevB.97.144304},
  url = {https://link.aps.org/doi/10.1103/PhysRevB.97.144304}
}

@article{Riddel2019,
  title = {Out-of-time ordered correlators and entanglement growth in the random-field XX spin chain},
  author = {Riddell, Jonathon and S\o{}rensen, Erik S.},
  journal = {Phys. Rev. B},
  volume = {99},
  issue = {5},
  pages = {054205},
  numpages = {13},
  year = {2019},
  month = {Feb},
  publisher = {American Physical Society},
  doi = {10.1103/PhysRevB.99.054205},
  url = {https://link.aps.org/doi/10.1103/PhysRevB.99.054205}
}

@article{Fazio2018,
  title={ Scrambling and entanglement spreading in long-range spin chains},
  author={Pappalardi, S and Russomanno, A and \v{Z}unkovi\v{c}, B and Iemini, F and Silva, A and Fazio, R},
  journal={arXiv:1806.00022},
  year={2018}
}

@article{XuPRL2020,
  title = {Does Scrambling Equal Chaos?},
  author = {Xu, Tianrui and Scaffidi, Thomas and Cao, Xiangyu},
  journal = {Phys. Rev. Lett.},
  volume = {124},
  issue = {14},
  pages = {140602},
  numpages = {7},
  year = {2020},
  month = {Apr},
  publisher = {American Physical Society},
  doi = {10.1103/PhysRevLett.124.140602},
  url = {https://link.aps.org/doi/10.1103/PhysRevLett.124.140602}
}

@article{PilatowskiPRE2020,
  title = {Positive quantum {L}yapunov exponents in experimental systems with a regular classical limit},
  author = {Pilatowsky-Cameo, Sa\'ul and Ch\'avez-Carlos, Jorge and Bastarrachea-Magnani, Miguel A. and Str\'ansk\'y, Pavel and Lerma-Hern\'andez, Sergio and Santos, Lea F. and Hirsch, Jorge G.},
  journal = {Phys. Rev. E},
  volume = {101},
  issue = {1},
  pages = {010202},
  numpages = {7},
  year = {2020},
  month = {Jan},
  publisher = {American Physical Society},
  doi = {10.1103/PhysRevE.101.010202},
  url = {https://link.aps.org/doi/10.1103/PhysRevE.101.010202}
}

@article{HummelPRL2019,
  title = {Reversible Quantum Information Spreading in Many-Body Systems near Criticality},
  author = {Hummel, Quirin and Geiger, Benjamin and Urbina, Juan Diego and Richter, Klaus},
  journal = {Phys. Rev. Lett.},
  volume = {123},
  issue = {16},
  pages = {160401},
  numpages = {7},
  year = {2019},
  month = {Oct},
  publisher = {American Physical Society},
  doi = {10.1103/PhysRevLett.123.160401},
  url = {https://link.aps.org/doi/10.1103/PhysRevLett.123.160401}
}

@article{Bhattacharyya2022,
	Author = {Bhattacharyya, Arpan and Chemissany, Wissam and Haque, S. Shajidul and Yan, Bin},
	Da = {2022/01/29},
	Doi = {10.1140/epjc/s10052-022-10035-3},
	Id = {Bhattacharyya2022},
	Isbn = {1434-6052},
	Journal = {The European Physical Journal C},
	Number = {1},
	Pages = {87},
	Title = {Towards the web of quantum chaos diagnostics},
	Ty = {JOUR},
	Url = {https://doi.org/10.1140/epjc/s10052-022-10035-3},
	Volume = {82},
	Year = {2022}}

@article{hashimoto2020exponential,
  title={Exponential growth of out-of-time-order correlator without chaos: inverted harmonic oscillator},
  author={Hashimoto, Koji and Huh, Kyoung-Bum and Kim, Keun-Young and Watanabe, Ryota},
  journal={Journal of High Energy Physics},
  volume={11},
  pages={068},
  year={2020},
  url={https://link.springer.com/article/10.1007/JHEP11(2020)068},
  publisher={Springer}
}

@Article{Bhattacharyya2021multi,
	title={{The multi-faceted inverted harmonic oscillator: Chaos and complexity}},
	author={Arpan Bhattacharyya and Wissam Chemissany and S. Shajidul Haque and Jeff Murugan and Bin Yan},
	journal={SciPost Phys. Core},
	volume={4},
	pages={002},
	year={2021},
	publisher={SciPost},
	doi={10.21468/SciPostPhysCore.4.1.002},
	url={https://scipost.org/10.21468/SciPostPhysCore.4.1.002},
}

@article{kidd2021saddle,
  title = {Saddle-point scrambling without thermalization},
  author = {Kidd, R. A. and Safavi-Naini, A. and Corney, J. F.},
  journal = {Phys. Rev. A},
  volume = {103},
  issue = {3},
  pages = {033304},
  numpages = {7},
  year = {2021},
  month = {Mar},
  publisher = {American Physical Society},
  doi = {10.1103/PhysRevA.103.033304},
  url = {https://link.aps.org/doi/10.1103/PhysRevA.103.033304}
}

@article{notenson2023classical,
  title = {Classical approach to equilibrium of out-of-time ordered correlators in mixed systems},
  author = {Notenson, Tom\'as and Garc\'{\i}a-Mata, Ignacio and Roncaglia, Augusto J. and Wisniacki, Diego A.},
  journal = {Phys. Rev. E},
  volume = {107},
  issue = {6},
  pages = {064207},
  numpages = {8},
  year = {2023},
  month = {Jun},
  publisher = {American Physical Society},
  doi = {10.1103/PhysRevE.107.064207},
  url = {https://link.aps.org/doi/10.1103/PhysRevE.107.064207}
}

@article{pollicott1985rate,
  title={On the rate of mixing of {A}xiom {A} flows},
  author={Pollicott, Mark},
  journal={Inventiones mathematicae},
  volume={81},
  number={3},
  pages={413--426},
  year={1985},
  url={https://link.springer.com/article/10.1007/BF01388579#citeas},
  publisher={Springer}
}

@article{ruelle1986,
  title={Locating resonances for {A}xiom {A} dynamical systems},
  author={Ruelle, David},
  journal={J. Stat. Phys.},
  volume={44},
  number={3-4},
  pages={281--292},
  year={1986},
  url={https://link.springer.com/article/10.1007/BF01011300},
  publisher={Springer}
}

@article{ruelle1987resonances,
  title={Resonances for {A}xiom {A} flows},
  author={Ruelle, David},
  journal={J. Differential Geom},
  volume={25},
  number={1},
  pages={99--116},
  url={https://projecteuclid.org/journals/journal-of-differential-geometry/volume-25/issue-1/Resonances-for-Axiom-bf-A-flows/10.4310/jdg/1214440726.full},
  year={1987}
}

@article{Prosen2002Ruelle,
  author  = {Prosen, Toma\v{z}},
  title   = {Ruelle resonances in quantum many-body dynamics},
  journal = {Journal of Physics A: Mathematical and General},
  volume  = {35},
  number  = {44},
  pages   = {L737--L744},
  year    = {2002},
  doi     = {10.1088/0305-4470/35/44/101}
}

@article{Prosen2004Ruelle,
  author  = {Prosen, Toma\v{z}},
  title   = {Ruelle resonances in kicked quantum spin chain},
  journal = {Physica D: Nonlinear Phenomena},
  volume  = {187},
  number  = {1-4},
  pages   = {244--255},
  year    = {2004},
  doi     = {10.1016/j.physd.2003.09.017}
}

@article{Prosen2007Chaos,
  author  = {Prosen, Toma\v{z}},
  title   = {Chaos and complexity of quantum motion},
  journal = {Journal of Physics A: Mathematical and Theoretical},
  volume  = {40},
  number  = {28},
  pages   = {7881--7892},
  year    = {2007},
  doi     = {10.1088/1751-8113/40/28/S11}
}

@article{znidaric2024momentum,
  title = {Momentum-dependent quantum Ruelle-Pollicott resonances in translationally invariant many-body systems},
  author = {\ifmmode \check{Z}\else \v{Z}\fi{}nidari\ifmmode \check{c}\else \v{c}\fi{}, Marko},
  journal = {Phys. Rev. E},
  volume = {110},
  issue = {5},
  pages = {054204},
  numpages = {13},
  year = {2024},
  month = {Nov},
  publisher = {American Physical Society},
  doi = {10.1103/PhysRevE.110.054204},
  url = {https://link.aps.org/doi/10.1103/PhysRevE.110.054204}
}

@article{Jacoby2025,
  author  = {Jacoby, Jacob A. and Huse, David A. and Gopalakrishnan, Sarang},
  title   = {Spectral gaps of local quantum channels in the weak-dissipation limit},
  journal = {Physical Review B},
  volume  = {111},
  pages   = {104303},
  year    = {2025},
  doi     = {10.1103/PhysRevB.111.104303}
}

@article{Zhang2024RPR,
  author        = {Zhang, Cheng and Nie, Lin and von Keyserlingk, C. W.},
  title         = {Thermalization rates and quantum Ruelle-Pollicott resonances: Insights from operator hydrodynamics},
  journal       = {arXiv preprint arXiv:2409.17251},
  year          = {2024},
  eprint        = {2409.17251},
  archivePrefix = {arXiv},
  primaryClass  = {cond-mat.stat-mech}
}

@article{Yoshimura2025,
  author  = {Yoshimura, Takato and S\'a, Leonardo},
  title   = {Theory of irreversibility in quantum many-body systems},
  journal = {Physical Review E},
  volume  = {111},
  pages   = {064135},
  year    = {2025},
  doi     = {10.1103/PhysRevE.111.064135}
}

@article{Duarte2026Ruelle,
  title = {Ruelle-Pollicott decay of out-of-time-order correlators in many-body systems},
  author = {Duarte, Jer\'onimo and Garc\'{\i}a-Mata, Ignacio and Wisniacki, Diego A.},
  journal = {Phys. Rev. E},
  volume = {113},
  issue = {2},
  pages = {024209},
  numpages = {8},
  year = {2026},
  month = {Feb},
  publisher = {American Physical Society},
  doi = {10.1103/5rlf-93y8},
  url = {https://link.aps.org/doi/10.1103/5rlf-93y8}
}

@article{RobertsYoshida2017,
  author  = {Roberts, Daniel A. and Yoshida, Beni},
  title   = {Chaos and complexity by design},
  journal = {Journal of High Energy Physics},
  volume  = {2017},
  number  = {4},
  pages   = {121},
  year    = {2017},
  doi     = {10.1007/JHEP04(2017)121},
  eprint  = {1610.04903},
  archivePrefix = {arXiv},
  primaryClass  = {hep-th}
}

@article{Kim2014,
  title={Local integrals of motion and the logarithmic lightcone in many-body localized systems},
  author={Kim, Isaac H and Chandran, Anushya and Abanin, Dmitry A},
  journal={arXiv preprint arXiv:1412.3073},
  url={https://arxiv.org/abs/1412.3073},
year={2014}
}

@article{chen2016universal,
  title={Universal logarithmic scrambling in many body localization},
  author={Chen, Yu},
  journal={arXiv preprint arXiv:1608.02765},
  year={2016}
}

@article{smith2019logarithmic,
  title = {Logarithmic Spreading of Out-of-Time-Ordered Correlators without Many-Body Localization},
  author = {Smith, Adam and Knolle, Johannes and Moessner, Roderich and Kovrizhin, Dmitry L.},
  journal = {Phys. Rev. Lett.},
  volume = {123},
  issue = {8},
  pages = {086602},
  numpages = {6},
  year = {2019},
  month = {Aug},
  publisher = {American Physical Society},
  doi = {10.1103/PhysRevLett.123.086602},
  url = {https://link.aps.org/doi/10.1103/PhysRevLett.123.086602}
}

@article{fan2017out,
  title={Out-of-time-order correlation for many-body localization},
  author={Fan, Ruihua and Zhang, Pengfei and Shen, Huitao and Zhai, Hui},
  journal={Science bulletin},
  volume={62},
  number={10},
  pages={707--711},
  year={2017},
  url={https://www.sciencedirect.com/science/article/pii/S2095927317301925},
  publisher={Elsevier}
}

@article{Huang2016,
	doi = {10.1002/andp.201600318},
	url = {https://doi.org/10.1002%2Fandp.201600318},
	year = 2016,
	month = {dec},
	publisher = {Wiley},
	volume = {529},
	number = {7},
	pages = {1600318},
	author = {Yichen Huang and Yong-Liang Zhang and Xie Chen},
	title = {Out-of-time-ordered correlators in many-body localized systems},
	journal = {Annalen der Physik}
}

@article{chen2017out,
  title={Out-of-time-order correlations in many-body localized and thermal phases},
  author={Chen, Xiao and Zhou, Tianci and Huse, David A and Fradkin, Eduardo},
  journal={Annalen der Physik},
  volume={529},
  number={7},
  pages={1600332},
  year={2017},
  url={https://doi.org/10.1002/andp.201600332},
  publisher={Wiley Online Library}
}

@article{Garttner2017,
author = {G{\"a}rttner, Martin and Bohnet, Justin G and Safavi-Naini, Arghavan and Wall, Michael L and Bollinger, John J and Rey, Ana Maria},
title = {{Measuring out-of-time-order correlations and multiple quantum spectra in a trapped-ion quantum magnet}},
journal = {Nature Physics},
year = {2017},
volume = {13},
number = {8},
pages = {781--786},
url={https://www.nature.com/articles/nphys4119},
month = aug
}

@article{joshi2020quantum,
  title = {Quantum Information Scrambling in a Trapped-Ion Quantum Simulator with Tunable Range Interactions},
  author = {Joshi, Manoj K. and Elben, Andreas and Vermersch, Beno\^{\i}t and Brydges, Tiff and Maier, Christine and Zoller, Peter and Blatt, Rainer and Roos, Christian F.},
  journal = {Phys. Rev. Lett.},
  volume = {124},
  issue = {24},
  pages = {240505},
  numpages = {6},
  year = {2020},
  month = {Jun},
  publisher = {American Physical Society},
  doi = {10.1103/PhysRevLett.124.240505},
  url = {https://link.aps.org/doi/10.1103/PhysRevLett.124.240505}
}

@article{green2022experimental,
  title = {Experimental Measurement of Out-of-Time-Ordered Correlators at Finite Temperature},
  author = {Green, Alaina M. and Elben, A. and Alderete, C. Huerta and Joshi, Lata Kh and Nguyen, Nhung H. and Zache, Torsten V. and Zhu, Yingyue and Sundar, Bhuvanesh and Linke, Norbert M.},
  journal = {Phys. Rev. Lett.},
  volume = {128},
  issue = {14},
  pages = {140601},
  numpages = {6},
  year = {2022},
  month = {Apr},
  publisher = {American Physical Society},
  doi = {10.1103/PhysRevLett.128.140601},
  url = {https://link.aps.org/doi/10.1103/PhysRevLett.128.140601}
}

@article{Li2017,
  title = {Measuring Out-of-Time-Order Correlators on a Nuclear Magnetic Resonance Quantum Simulator},
  author = {Li, Jun and Fan, Ruihua and Wang, Hengyan and Ye, Bingtian and Zeng, Bei and Zhai, Hui and Peng, Xinhua and Du, Jiangfeng},
  journal = {Phys. Rev. X},
  volume = {7},
  issue = {3},
  pages = {031011},
  numpages = {12},
  year = {2017},
  month = {Jul},
  publisher = {American Physical Society},
  doi = {10.1103/PhysRevX.7.031011},
  url = {https://link.aps.org/doi/10.1103/PhysRevX.7.031011}
}

@article{wei2018exploring,
  title = {Exploring Localization in Nuclear Spin Chains},
  author = {Wei, Ken Xuan and Ramanathan, Chandrasekhar and Cappellaro, Paola},
  journal = {Phys. Rev. Lett.},
  volume = {120},
  issue = {7},
  pages = {070501},
  numpages = {6},
  year = {2018},
  month = {Feb},
  publisher = {American Physical Society},
  doi = {10.1103/PhysRevLett.120.070501},
  url = {https://link.aps.org/doi/10.1103/PhysRevLett.120.070501}
}

@article{Nie2020experimental,
  title = {Experimental Observation of Equilibrium and Dynamical Quantum Phase Transitions via Out-of-Time-Ordered Correlators},
  author = {Nie, Xinfang and Wei, Bo-Bo and Chen, Xi and Zhang, Ze and Zhao, Xiuzhu and Qiu, Chudan and Tian, Yu and Ji, Yunlan and Xin, Tao and Lu, Dawei and Li, Jun},
  journal = {Phys. Rev. Lett.},
  volume = {124},
  issue = {25},
  pages = {250601},
  numpages = {6},
  year = {2020},
  month = {Jun},
  publisher = {American Physical Society},
  doi = {10.1103/PhysRevLett.124.250601},
  url = {https://link.aps.org/doi/10.1103/PhysRevLett.124.250601}
}

@article{MohamadPRR2020,
  title = {Sensitivity of quantum information to environment perturbations measured with a nonlocal out-of-time-order correlation function},
  author = {Niknam, Mohamad and Santos, Lea F. and Cory, David G.},
  journal = {Phys. Rev. Research},
  volume = {2},
  issue = {1},
  pages = {013200},
  numpages = {13},
  year = {2020},
  month = {Feb},
  publisher = {American Physical Society},
  doi = {10.1103/PhysRevResearch.2.013200},
  url = {https://link.aps.org/doi/10.1103/PhysRevResearch.2.013200}
}

@article{zhou2023operator,
  title={Operator growth from global out-of-time-order correlators},
  author={Zhou, Tianci and Swingle, Brian},
  journal={Nature communications},
  volume={14},
  number={1},
  pages={3411},
  year={2023},
  publisher={Nature Publishing Group UK London}
}

@article{braumuller2022probing,
  title={Probing quantum information propagation with out-of-time-ordered correlators},
  author={Braum{\"u}ller, Jochen and Karamlou, Amir H and Yanay, Yariv and Kannan, Bharath and Kim, David and Kjaergaard, Morten and Melville, Alexander and Niedzielski, Bethany M and Sung, Youngkyu and Veps{\"a}l{\"a}inen, Antti and others},
  journal={Nature Physics},
  volume={18},
  number={2},
  pages={172--178},
  year={2022},
  publisher={Nature Publishing Group UK London}
}

@article{Zhao2022probing,
  title = {Probing Operator Spreading via Floquet Engineering in a Superconducting Circuit},
  author = {Zhao, S. K. and Ge, Zi-Yong and Xiang, Zhongcheng and Xue, G. M. and Yan, H. S. and Wang, Z. T. and Wang, Zhan and Xu, H. K. and Su, F. F. and Yang, Z. H. and Zhang, He and Zhang, Yu-Ran and Guo, Xue-Yi and Xu, Kai and Tian, Ye and Yu, H. F. and Zheng, D. N. and Fan, Heng and Zhao, S. P.},
  journal = {Phys. Rev. Lett.},
  volume = {129},
  issue = {16},
  pages = {160602},
  numpages = {7},
  year = {2022},
  month = {Oct},
  publisher = {American Physical Society},
  doi = {10.1103/PhysRevLett.129.160602},
  url = {https://link.aps.org/doi/10.1103/PhysRevLett.129.160602}
}

@article{google2025Observation,
	Author = {Abanin, Dmitry A. and Acharya, Rajeev and Aghababaie-Beni, Laleh and Aigeldinger, Georg and Ajoy, Ashok and Alcaraz, Ross and Aleiner, Igor and Andersen, Trond I. and Ansmann, Markus and Arute, Frank and Arya, Kunal and Asfaw, Abraham and Astrakhantsev, Nikita and Atalaya, Juan and Babbush, Ryan and Bacon, Dave and Ballard, Brian and Bardin, Joseph C. and Bengs, Christian and Bengtsson, Andreas and Bilmes, Alexander and Boixo, Sergio and Bortoli, Gina and Bourassa, Alexandre and Bovaird, Jenna and Bowers, Dylan and Brill, Leon and Broughton, Michael and Browne, David A. and Buchea, Brett and Buckley, Bob B. and Buell, David A. and Burger, Tim and Burkett, Brian and Bushnell, Nicholas and Cabrera, Anthony and Campero, Juan and Chang, Hung-Shen and Chen, Yu and Chen, Zijun and Chiaro, Ben and Chih, Liang-Ying and Chik, Desmond and Chou, Charina and Claes, Jahan and Cleland, Agnetta Y. and Cogan, Josh and Cohen, Saul and Collins, Roberto and Conner, Paul and Courtney, William and Crook, Alexander L. and Curtin, Ben and Das, Sayan and De Lorenzo, Laura and Debroy, Dripto M. and Demura, Sean and Devoret, Michel and Di Paolo, Agustin and Donohoe, Paul and Drozdov, Ilya and Dunsworth, Andrew and Earle, Clint and Eickbusch, Alec and Elbag, Aviv Moshe and Elzouka, Mahmoud and Erickson, Catherine and Faoro, Lara and Farhi, Edward and Ferreira, Vinicius S. and Burgos, Leslie Flores and Forati, Ebrahim and Fowler, Austin G. and Foxen, Brooks and Ganjam, Suhas and Garcia, Gonzalo and Gasca, Robert and Genois, {\'E}lie and Giang, William and Gidney, Craig and Gilboa, Dar and Gosula, Raja and Dau, Alejandro Grajales and Graumann, Dietrich and Greene, Alex and Gross, Jonathan A. and Gu, Hanfeng and Habegger, Steve and Hall, John and Hamamura, Ikko and Hamilton, Michael C. and Hansen, Monica and Harrigan, Matthew P. and Harrington, Sean D. and Heslin, Stephen and Heu, Paula and Higgott, Oscar and Hill, Gordon and Hilton, Jeremy and Hong, Sabrina and Huang, Hsin-Yuan and Huff, Ashley and Huggins, William J. and Ioffe, Lev B. and Isakov, Sergei V. and Iveland, Justin and Jeffrey, Evan and Jiang, Zhang and Jin, Xiaoxuan and Jones, Cody and Jordan, Stephen and Joshi, Chaitali and Juhas, Pavol and Kabel, Andreas and Kafri, Dvir and Kang, Hui and Karamlou, Amir H. and Kechedzhi, Kostyantyn and Kelly, Julian and Khaire, Trupti and Khattar, Tanuj and Khezri, Mostafa and Kim, Seon and King, Robbie and Klimov, Paul V. and Klots, Andrey R. and Kobrin, Bryce and Korotkov, Alexander N. and Kostritsa, Fedor and Kothari, Robin and Kreikebaum, John Mark and Kurilovich, Vladislav D. and Kyoseva, Elica and Landhuis, David and Lange-Dei, Tiano and Langley, Brandon W. and Laptev, Pavel and Lau, Kim-Ming and Le Guevel, Lo{\"\i}ck and Ledford, Justin and Lee, Joonho and Lee, Kenny and Lensky, Yuri D. and Leon, Shannon and Lester, Brian J. and Li, Wing Yan and Lill, Alexander T. and Liu, Wayne and Livingston, William P. and Locharla, Aditya and Lucero, Erik and Lundahl, Daniel and Lunt, Aaron and Madhuk, Sid and Malone, Fionn D. and Maloney, Ashley and Mandr{\`a}, Salvatore and Manyika, James M. and Martin, Leigh S. and Martin, Orion and Martin, Steven and Matias, Yossi and Maxfield, Cameron and McClean, Jarrod R. and McEwen, Matt and Meeks, Seneca and Megrant, Anthony and Mi, Xiao and Miao, Kevin C. and Mieszala, Amanda and Minev, Zlatko and Molavi, Reza and Molina, Sebastian and Montazeri, Shirin and Morvan, Alexis and Movassagh, Ramis and Mruczkiewicz, Wojciech and Naaman, Ofer and Neeley, Matthew and Neill, Charles and Nersisyan, Ani and Neven, Hartmut and Newman, Michael and Ng, Jiun How and Nguyen, Anthony and Nguyen, Murray and Ni, Chia-Hung and Niu, Murphy Yuezhen and Oas, Logan and O'Brien, Thomas E. and Oliver, William D. and Opremcak, Alex and Ottosson, Kristoffer and Petukhov, Andre and Pizzuto, Alex and Platt, John and Potter, Rebecca and Pritchard, Orion and Pryadko, Leonid P. and Quintana, Chris and Ramachandran, Ganesh and Ramanathan, Chandrasekhar and Reagor, Matthew J. and Redding, John and Rhodes, David M. and Roberts, Gabrielle and Rosenberg, Eliott and Rosenfeld, Emma and Roushan, Pedram and Rubin, Nicholas C. and Saei, Negar and Sank, Daniel and Sankaragomathi, Kannan and Satzinger, Kevin J. and Schmidhuber, Alexander and Schurkus, Henry F. and Schuster, Christopher and Schuster, Thomas and Shearn, Michael J. and Shorter, Aaron and Shutty, Noah and Shvarts, Vladimir and Sivak, Volodymyr and Skruzny, Jindra and Small, Spencer and Smelyanskiy, Vadim and Smith, W. Clarke and Somma, Rolando D. and Springer, Sofia and Sterling, George and Strain, Doug and Suchard, Jordan and Suchsland, Philippe and Szasz, Aaron and Sztein, Alex and Thor, Douglas and Tomita, Eifu and Torres, Alfredo and Torunbalci, M. Mert and Vaishnav, Abeer and Vargas, Justin and Vdovichev, Sergey and Vidal, Guifre and Villalonga, Benjamin and Heidweiller, Catherine Vollgraff and Waltman, Steven and Wang, Shannon X. and Ware, Brayden and Weber, Kate and Weidel, Travis and Westerhout, Tom and White, Theodore and Wong, Kristi and Woo, Bryan W. K. and Xing, Cheng and Yao, Z. Jamie and Yeh, Ping and Ying, Bicheng and Yoo, Juhwan and Yosri, Noureldin and Young, Grayson and Zalcman, Adam and Zhang, Chongwei and Zhang, Yaxing and Zhu, Ningfeng and Zobrist, Nicholas and Google Quantum AI and Collaborators},
	Da = {2025/10/01},
	Doi = {10.1038/s41586-025-09526-6},
	Id = {Abanin2025},
	Isbn = {1476-4687},
	Journal = {Nature},
	Number = {8086},
	Pages = {825--830},
	Title = {Observation of constructive interference at the edge of quantum ergodicity},
	Ty = {JOUR},
	Url = {https://doi.org/10.1038/s41586-025-09526-6},
	Volume = {646},
	Year = {2025}}

@article{Pappalardi2024,
  author  = {Pappalardi, Silvia and others},
  title   = {Higher-order out-of-time-ordered correlators and refined diagnostics of scrambling},
  journal = {arXiv preprint arXiv:2411.12050},
  year    = {2024},
  eprint  = {2411.12050},
  archivePrefix = {arXiv},
  primaryClass  = {cond-mat.stat-mech}
}

@article{GoogleOTOC2025,
  author  = {Google Quantum AI and collaborators},
  title   = {Observation of constructive interference at the edge of quantum ergodicity},
  journal = {Nature},
  year    = {2025},
  doi     = {10.1038/s41586-025-09526-6}
}

@article{hochbruck1997krylov,
  title={On Krylov subspace approximations to the matrix exponential operator},
  author={Hochbruck, Marlis and Lubich, Christian},
  journal={SIAM Journal on Numerical Analysis},
  volume={34},
  number={5},
  pages={1911--1925},
  year={1997},
  publisher={SIAM}
}

@book{parlett1998symmetric,
  title={The symmetric eigenvalue problem},
  author={Parlett, Beresford N},
  year={1998},
  publisher={SIAM}
}

@article{Rabinovici2021,
    title = {Operator complexity: A journey to the edge of Krylov space},
    author = {Rabinovici, E. and Sanchez-Garrido, A. and Shir, R. and Sonner, J.},
    journal = {J. High Energy Phys.},
    volume = {2021},
    pages = {62},
    year = {2021},
    doi = {10.1007/JHEP06(2021)062}
}

@article{Balasubramanian2022,
    title = {Quantum chaos and the complexity of spread of states},
    author = {Balasubramanian, Vijay and Caputa, Pawel and Mag{\'a}n, Javier M. and Wu, Qingyue},
    journal = {Phys. Rev. D},
    volume = {106},
    issue = {4},
    pages = {046007},
    year = {2022},
    doi = {10.1103/PhysRevD.106.046007}
}

@article{Rabinovici2022,
    title = {Krylov complexity from integrability to chaos},
    author = {Rabinovici, E. and Sanchez-Garrido, A. and Shir, R. and Sonner, J.},
    journal = {J. High Energy Phys.},
    volume = {2022},
    pages = {151},
    year = {2022},
    doi = {10.1007/JHEP07(2022)151}
}

@article{berry1977,
    title={Regular and irregular semiclassical wavefunctions},
    author={Berry, Michael V},
    journal={Journal of Physics A: Mathematical and General},
    volume={10},
    number={12},
    pages={2083},
    year={1977},
    publisher={IOP Publishing}
}

@article{dymarsky2021,
    title={Krylov complexity in quantum field theory},
    author={Dymarsky, Anatoly and Smolkin, Michael},
    journal={Physical Review D},
    volume={104},
    number={8},
    pages={L081702},
    year={2021},
    publisher={APS}
}

@article{Espanol2023,
    title = {Assessing the saturation of Krylov complexity as a measure of chaos},
    author = {Espa{\~n}ol, Bernardo L. and Wisniacki, Diego A.},
    journal = {Phys. Rev. E},
    volume = {107},
    issue = {2},
    pages = {024217},
    year = {2023},
    doi = {10.1103/PhysRevE.107.024217}
}

@article{Scialchi2024,
    title = {Integrability-to-chaos transition through the Krylov approach for state evolution},
    author = {Scialchi, Gaston F. and Roncaglia, Augusto J. and Wisniacki, Diego A.},
    journal = {Phys. Rev. E},
    volume = {109},
    issue = {5},
    pages = {054209},
    year = {2024},
    doi = {10.1103/PhysRevE.109.054209}
}

@article{nation2021,
    title={Quantum chaos in the Krylov basis},
    author={Nation, Andrew and Sannomiya, Nobuyuki and Yoshida, Haruki},
    journal={Physical Review Research},
    volume={3},
    number={4},
    pages={043027},
    year={2021},
    publisher={APS}
}

@article{caputa2022,
    title={Krylov complexity in conformal field theory},
    author={Caputa, Pawel and Liu, Shouvik and Miyaji, Masamichi and Takayanagi, Tadashi},
    journal={Physical Review D},
    volume={106},
    number={4},
    pages={046017},
    year={2022},
    publisher={APS}
}

@article{pg2025dependence,
  title={Dependence of Krylov complexity saturation on the initial operator and state},
  author={Pg, Sreeram and Kannan, J Bharathi and Modak, Ranjan and Aravinda, S},
  journal={Physical Review E},
  volume={112},
  number={3},
  pages={L032203},
  year={2025},
  publisher={APS}
}

@article{Suchsland2025,
    title = {Krylov complexity and Trotter transitions in unitary circuit dynamics},
    author = {Suchsland, P. and Moessner, R. and Claeys, P. W.},
    journal = {Phys. Rev. B},
    volume = {111},
    issue = {1},
    pages = {014309},
    year = {2025},
    doi = {10.1103/PhysRevB.111.014309}
}

@article{Scialchi2025,
    title = {Exploring quantum ergodicity of unitary evolution through the Krylov approach},
    author = {Scialchi, Gaston F. and Roncaglia, Augusto J. and Pineda, Carlos and Wisniacki, Diego A.},
    journal = {Phys. Rev. E},
    volume = {111},
    issue = {1},
    pages = {014220},
    year = {2025},
    doi = {10.1103/PhysRevE.111.014220}
}

@article{craps2025multiseed,
  title={Multiseed krylov complexity},
  author={Craps, Ben and Evnin, Oleg and Pascuzzi, Gabriele},
  journal={Physical Review Letters},
  volume={134},
  number={5},
  pages={050402},
  year={2025},
  publisher={APS}
}

\end{document}